\documentclass[times]{elsart}

\usepackage{amsmath} 
\usepackage{amssymb} 
\usepackage{amsfonts}
\usepackage{graphicx,color}                 
\usepackage{subfigure}
\usepackage{rotating}
\usepackage{multirow}
\begin{document}

\begin{frontmatter}
\title{Ensemble Kalman filter with the unscented transform}

\author[maths,man]{X. Luo\corauthref{lxd}}
\corauth[lxd]{Corresponding author.}
\ead{luox@maths.ox.ac.uk}
\author[maths]{, I.M. Moroz}

\address[maths]{Mathematical Institute, 24-29 St Giles', Oxford, UK, OX1 3LB}
\address[man]{Oxford-Man Institute of Quantitative Finance, Blue Boar Court, 9 Alfred Street, Oxford, UK, OX1 4EH}

\begin{abstract}
A modification scheme to the ensemble Kalman filter (EnKF) is introduced based on the concept of the unscented transform (Julier et al., 2000; Julier and Uhlmann, 2004), which therefore will be called the ensemble unscented Kalman filter (EnUKF) in this work. When the error distribution of the analysis is symmetric (not necessarily Gaussian), it can be shown that, compared to the ordinary EnKF, the EnUKF has more accurate estimations of the ensemble mean and covariance of the background by examining the multidimensional Taylor series expansion term by term. This implies that, the EnUKF may have better performance in state estimation than the ordinary EnKF in the sense that the deviations from the true states are smaller. For verification, some numerical experiments are conducted on a $40$-dimensional system due to Lorenz and Emanuel (Lorenz and Emanuel, 1998). Simulation results support our argument.

\end{abstract}

\end{frontmatter}

\maketitle

\section{Introduction}

The Kalman filter (KF) is a recursive data processing algorithm \cite{Maybeck-stochastic}. It optimally estimates the states of linear stochastic systems that are driven by Gaussian noise, and are observed through linear observation operators, which possibly also suffer from additive Gaussian errors. However, if there exists nonlinearity from either the dynamical systems or the observation operators, or, if neither the dynamical noise nor the observational noise follows any Gaussian distribution, then the Kalman filter becomes suboptimal. To tackle the problems of nonlinearity and non-Gaussianity, there are some strategies one may employ. For example, to handle the problem of nonlinearity, one may expand the nonlinear function in a Taylor series locally and keep the expansion terms only up to second order. This leads to the extended Kalman filter (EKF) (e.g., \cite{Evensen-using}). To deal with the problem of non-Gaussianity, one may specify a Gaussian mixture model (GMM) to approximate the underlying probability density function (pdf), such that the KF algorithm is applicable to the individual distributions of the GMM \cite{Smith-cluster}. More generally, one may adopt the sequential Monte Carlo method (also known as the particle filter, e.g., \cite{vanLeeuwen-variance}), which utilizes the empirical pdf obtained from a number of particles to represent the true pdf, wherein the problems of both nonlinearity and non-Gaussianity are taken into account during the pdf approximation.

For practical large-scale problems like weather forecasting, the computational cost is another issue of great concern. In such circumstances, direct application of the KF or EKF scheme is prohibitive because of the computational cost of evolving the full covariance matrices forward. While for the particle filter, because of its slow convergence rate, the required number of the particles for proper approximations may be well above many thousands for even low dimensional nonlinear systems.

For the sake of computational efficiency, the ensemble Kalman filter (EnKF) was proposed \cite{Evensen-sequential}. It is essentially a Monte Carlo implementation of the Kalman filter.  At the beginning of each assimilation cycle, it is assumed that one has an ensemble of the system states, called the background ensemble, which can usually be obtained from the previous assimilation cycle. Then, given additional information from the observations, one applies the KF scheme to update each individual member of the background ensemble. To do this, the mean and error covariance of the background are approximated by the sample mean and covariance of the ensemble. After the updates, one will obtain a new set of the system states, called the analysis ensemble in this work, which will then be used to estimate the true mean and covariance of the underlying system state. By propagating the ensemble members of the analysis forward through the system model, one obtains a new ensemble of the background for the next assimilation cycle. Therefore, by using the ensembles to approximate the true statistics (e.g., the mean and covariance) of the system states, the computational cost can be significantly reduced. Moreover, in the EnKF there is no need to linearize the system model as in the EKF. Instead, the nonlinear problem becomes the one of how to approximate the statistics of the pdf of a Gaussian distribution which is transformed by a nonlinear function. This point of view leads to the idea of the unscented transform (UT) \cite{Julier2000,Julier2004}, as will be introduced later.

Depending on whether to perturb the observations or not, the EnKF can be classified into two types: stochastic and deterministic  \cite{Kalnay-4dvar,Tippett-ensemble}. The stochastic EnKF uses the observations and the corresponding covariance matrix to produce an ensemble of the disturbed observations, which is then used to update the background ensemble. For examples, see  \cite{Burgers-analysis,Evensen-sequential,Evensen-assimilation,Houtekamer1998}. In contrast, the deterministic EnKF, often known as the ensemble square root filter (EnSRF hereafter), does not perturb the observations. Given a background ensemble, the EnSRF uses the observations to update the sample mean of the background, while the analysis ensemble is produced based on  the sample mean plus the perturbations. For examples, see \cite{Anderson-ensemble,Bishop-adaptive,Whitaker-ensemble}; also see the review of \cite{Tippett-ensemble}.  Apart from the aforementioned EnKFs, there are also some other variants, e.g., \cite{Beezley-morphing,Sakov2007,Wang-which,Zupanski-maximum}.

In this paper we will introduce a framework of the EnKF incorporating the concept of the unscented transform \cite{Julier2000,Julier2004}, which will therefore be called the ensemble unscented KF (EnUKF for short). The basic idea of the EnUKF is that, at the end of the filtering step of an EnKF, one adopts the unscented transform to generate a set of  carefully chosen system states, called the sigma points. The set of the sigma points is then treated as the analysis ensemble and propagated forward through the system model. At the next assimilation cycle, the mean and covariance of the background are estimated based on the propagations. In the Appendix, we show that, under the assumption of Gaussian errors, the accuracies of thus estimated mean and covariance are up to third order term in the Taylor series expansion (if available). Moreover, there are additional adjustable parameters in the framework which may be tuned to reduce the approximation error further. In contrast, for the ordinary EnKFs, we will show in the Appendix that the accuracies are normally up to second order. Therefore, by incorporating the unscented transform, one may improve the performance of the EnKF.

This paper is organized as follows. In section \ref{sec:background} we will review the general framework of the EnKF. Moreover we will also introduce the idea of the unscented transform. The accuracy analysis of the unscented transform is given in the Appendix. For comparison, we will also provide the accuracy analysis of the ordinary EnSRF. In section \ref{sec:EnUKF}, based on the concept of the unscented transform, we will propose a modification scheme to the EnKF. In section \ref{sec:results}, we will adopt the 40-dimensional model in \cite{Lorenz-optimal} as the testbed to examine the performance of the EnUKF and compare it to the ordinary EnSRF. Finally we will conclude the work in section \ref{sec:conclusions}.

\section{Background}\label{sec:background}

\subsection{The framework of the EnKF}

The EnKF is a Monte Carlo implementation of the Kalman filter, which inherits the framework of the KF by nature. For illustration, we consider the following scenario. Suppose that we have an $m$-dimensional discrete dynamical system
\begin{equation}\label{system model}
\mathbf{x}_{k+1}=\mathcal{M}_{k,k+1} \left( \mathbf{x}_{k}\right) + \mathbf{u}_k,
\end{equation}
where $\mathbf{x}_{k}$ denotes the $m$-dimensional system state at the instant $k$, $\mathcal{M}_{k,k+1}$ is the transition operator mapping  $\mathbf{x}_{k}$ to  $\mathbf{x}_{k+1}$, and $\mathbf{u}_k$ is the dynamical noise, which is assumed to be independent of the system state and observation noise (see below), with zero mean and covariance $\mathbf{Q}_k$. 

The above dynamical system is then measured by an observer $\mathcal{H}_{k}$ such that
\begin{equation}\label{observer}
\mathbf{y}_{k}=\mathcal{H}_{k} \left( \mathbf{x}_{k}\right) + \mathbf{v}_{k},
\end{equation}
where $\mathbf{y}_{k}$ is the $p$-dimensional observation at instant $k$, and $\mathbf{v}_{k}$ is the observation noise, which is independent of the system state and dynamical noise, with zero mean and covariance $\mathbf{R}_k$. For convenience of discussion, we also assume that there is an $n$-member ensemble of the analysis $\mathbf{X}^a_{k-1} = \{\mathbf{x}^a_{k-1,i}: i= 1, 2 \dotsb, n\}$ available at the end of the $(k-1)$-th assimilation cycle. For description, we split the framework of the EnKF into the propagation (or prediction) and filtering steps.

\subsubsection{The propagation step}
We define the following set
\[
\mathbf{X}^b_k=\left\{ \mathbf{x}^b_{k,i} : \mathbf{x}^b_{k,i} = \mathcal{M}_{k,k+1} \left( \mathbf{x}^a_{k-1,i} \right), i=1, 2, \dotsb, n \right\}
\]
as the predicted background ensemble at instant $k$, which is the set of propagations of the analysis ensemble $\mathbf{X}^a_{k-1}$ at the previous assimilation cycle. The sample mean $\hat{\mathbf{x}}_k^b$ and covariance $\hat{\mathbf{P}}_k^b$  can be evaluated according to the following unbiased estimators \footnote{In some works, e.g., \cite{Wang-which}, the authors may choose other estimators.}.
\begin{subequations} \label{background mean}
\begin{align}
&\hat{\mathbf{x}}_k^b = \frac{1}{n} \sum\limits_{i=1}^n \mathbf{x}^b_{k,i} \, , \\
&\hat{\mathbf{P}}_k^b = \frac{1}{n-1} \sum\limits_{i=1}^n \left( \mathbf{x}^b_{k,i} - \hat{\mathbf{x}}_k^b \right) \left( \mathbf{x}^b_{k,i} - \hat{\mathbf{x}}_k^b \right)^T + \mathbf{Q}_k \, .
\end{align}
\end{subequations}
In practice, the approximation covariance $\hat{\mathbf{P}}_k^b$ need not be calculated. Instead, it is customary to compute
\begin{equation} \label{background covariances}
\begin{split}
\hat{\mathbf{P}}_{xh}^k =&  \frac{1}{n-1} \sum\limits_{i=1}^n \left( \mathbf{x}^b_{k,i} - \hat{\mathbf{x}}_k^b \right) \left( \mathcal{H}_k \left ( \mathbf{x}^b_{k,i} \right )- \mathcal{H}_k \left ( \hat{\mathbf{x}}_k^b \right)\right ) ^T , \\
\hat{\mathbf{P}}_{hh}^k = & \frac{1}{n-1} \sum\limits_{i=1}^n \left( \mathcal{H}_k \left ( \mathbf{x}^b_{k,i} \right )- \mathcal{H}_k \left ( \hat{\mathbf{x}}_k^b \right)\right ) \left( \mathcal{H}_k \left ( \mathbf{x}^b_{k,i} \right )- \mathcal{H}_k \left ( \hat{\mathbf{x}}_k^b \right)\right ) ^T, \\
\end{split}
\end{equation}
which avoids the problem of linearizing $\mathcal{H}_k$ when it is nonlinear, although under such circumstance the ordinary KF algorithm becomes sub-optimal. The Kalman gain $\mathbf{K}_k$ is obtained through
\begin{equation} \label{Kalman gain}
\mathbf{K}_k = \hat{\mathbf{P}}_{xh}^k \left( \hat{\mathbf{P}}_{hh}^k + \mathbf{R}_k \right)^{-1},
\end{equation}
where $\mathbf{R}_k$ is the covariance of the observational error at step $k$. 

\subsubsection{The filtering step}
With additional information from the observations, one can update the background ensemble according to a certain analysis scheme. For example, for a stochastic EnKF, the analysis ensemble $\mathbf{X}_k^a=\left\{ \mathbf{x}_{k,i}^a : i=1, 2, \dotsb, n \right\}$ can be generated according to
\begin{equation} \label{stochastic analysis scheme}
\mathbf{x}_{k,i}^a = \mathbf{x}_{k,i}^b + \mathbf{K}_k \left( \mathbf{y}_{k,i} - \mathcal{H}_k \left( \mathbf{x}_{k,i}^b \right) \right), \text{for } i = 1, 2, \dotsb, n,
\end{equation}
where $ \mathbf{y}_{k,i}$ is an observation sample generated by the normal distribution with mean $\mathbf{y}_k$ and covariance $\mathbf{R}_k$. Correspondingly, the sample mean and covariance of the analysis ensemble can be calculated according to
\begin{subequations}
\begin{align}
\label{analysis mean} &\hat{\mathbf{x}}_k^a = \frac{1}{n} \sum\limits_{i=1}^n \mathbf{x}_{k,i}^a \, , \\
\label{analysis cov} &\hat{\mathbf{P}}_k^{a} =  \frac{1}{n-1} \sum\limits_{i=1}^n \left( \mathbf{x}_{k,i}^a - \hat{\mathbf{x}}_k^a \right) \left(  \mathbf{x}_{k,i}^a - \hat{\mathbf{x}}_k^a \right ) ^T .
\end{align}
\end{subequations}

For an EnSRF, for instance, the ensemble transform Kalman filter (ETKF) \cite{Bishop-adaptive,Wang-which}, the analysis ensemble is generated by adding perturbations to the ensemble mean. On one hand, the ensemble mean $\hat{\mathbf{x}}_k^a$ is obtained as follows
\begin{equation} \label{entf: ensemble mean}
\hat{\mathbf{x}}_k^a = \hat{ \mathbf{x}}_k^b + \mathbf{K}_k \left(  \mathbf{y}_k - \mathcal{H} \left(  \hat{ \mathbf{x}}_k^b \right) \right).
\end{equation}
On the other hand, let $\mathbf{S}_k^b$ be an $m \times n$ square root matrix of the background error covariance $\hat{\mathbf{P}}_k^b$ such that $\hat{\mathbf{P}}_k^b =\mathbf{S}_k^b \left( \mathbf{S}_k^b \right)^T$\footnote{In general $\mathbf{S}_k^b$ can be obtained through some numerical algorithm, e.g., Cholesky decomposition.}, then an $m \times n$ square root matrix $\mathbf{S}_k^a$ of the analysis error covariance $\hat{\mathbf{P}}_k^a$ can be updated from $\mathbf{S}_k^b$ according to
\begin{equation}
\mathbf{S}_k^a = \mathbf{S}_k^b \, \mathbf{T}_k ,
\end{equation}
where $\mathbf{T}_k$ is an $n \times n$  transformation matrix derived in \cite{Bishop-adaptive}. Thus the analysis error covariance $\hat{\mathbf{P}}_k^a$ is computed by
\begin{equation} \label{entf: ensemble cov}
\hat{\mathbf{P}}_k^a = \mathbf{S}_k^a \left(\mathbf{S}_k^a \right)^T .
\end{equation}

Moreover, given $\hat{\mathbf{x}}_k^a$ and $ \mathbf{S}_k^a$, the analysis ensemble $\mathbf{X}_k^a=\left\{ \mathbf{x}_{k,i}^a : i=1, 2, \dotsb, n \right\}$ is generated according to 
\begin{equation}
\mathbf{x}_{k,i}^a = \hat{\mathbf{x}}_k^a +\sqrt{n-1} \left( \mathbf{S}_k^a \right)_i, \, i=1, 2, \dotsb, n, 
\end{equation}
where $\left( \mathbf{S}_k^a \right)_i$ denotes the $i$-th column of the square root matrix $\mathbf{S}_k^a$. After the analysis ensemble is generated, one propagates it forward to obtain the background ensemble at the next instant and starts a new assimilation cycle.

\subsection{The unscented transform}

Given an $m$-dimensional Gaussian random variable $\mathbf{x}$ with mean $\bar{\mathbf{x}}$ and covariance $\mathbf{P}_{xx}$, one problem of interest is how to estimate the mean and covariance of the transformed variable $\mathbf{y} = \mathbf{f} \left(\mathbf{x} \right)$, where $\mathbf{f}$ is a nonlinear vector function.

In the ordinary EnKF, given a set of samples $\{\mathbf{x}_i, i=1,2,\dotsb,n \}$ of the random variable $\mathbf{x}$ , the mean and covariance of the transformed variable $\mathbf{y}$ are estimated according to Eq. (\ref{background mean}), i.e.,
\begin{subequations}
\begin{align}
\label{enkf mean} &\hat{\mathbf{y}} = \frac{1}{n} \sum\limits_{i=1}^n \mathbf{f} \left(\mathbf{x}_i\right)  \, , \\
 \label{enkf cov} &\hat{\mathbf{P}}_{yy} = \frac{1}{n-1} \sum\limits_{i=1}^n \left(\mathbf{f} \left(\mathbf{x}_i\right)  - \hat{\mathbf{y}} \right) \left( \mathbf{f} \left(\mathbf{x}_i\right)  - \hat{\mathbf{y}} \right ) ^T .
\end{align}
\end{subequations}

The unscented transform \cite{Julier2000,Julier2004} is a different
method for the estimation problem. Suppose that $\bar{\mathbf{x}}$ and
$\mathbf{P}_{xx}$ of the random vector $\mathbf{x}$ are unknown, but
the sample mean $\hat{\mathbf{x}}$ and covariance
$\hat{\mathbf{P}}_{xx}$  can be estimated 
from the set $\{\mathbf{x}_i, i=1,2,\dotsb,n \}$. Then a set of $2L+1$ state vectors
$\{ \mathcal{X}_i, i=0, 1, \dotsb, 2L \}$, called the sigma
points \cite{Julier2000,Julier2004}, can be generated according to
\begin{equation} \label{sigma points}
\begin{split}
& \mathcal{X}_0 = \hat{\mathbf{x}},\\
& \mathcal{X}_i = \hat{\mathbf{x}} + \left( \sqrt{(L+\lambda) \hat{\mathbf{P}}_{xx}} \right)_i, \, i=1, 2, \dotsb, L ,\\
& \mathcal{X}_i = \hat{\mathbf{x}} - \left( \sqrt{(L+\lambda) \hat{\mathbf{P}}_{xx}} \right)_{i-L}, \, i=L+1, L+2, \dotsb, 2L ,\\
\end{split}
\end{equation}
where $\left( \sqrt{(L+\lambda) \hat{\mathbf{P}}_{xx}} \right)_i$
denotes the $i$-th column of the square root matrix $
\sqrt{(L+\lambda) \hat{\mathbf{P}}_{xx}}$, and $\lambda$ is an
adjustable scaling parameter \cite{Julier2000,Julier2004}.

Moreover, in the unscented transform, it also allocates a set of weights $\{ W_i, i=0,1,\dotsb,2L \}$
\begin{equation} \label{ut weights}
\begin{split}
&W_0 = \frac{\lambda}{L+\lambda},\\
&W_i = \frac{1}{2\left(L+\lambda  \right)}, \, i=1,2,\dotsb, 2L, \\
\end{split}
\end{equation}
to the sigma points. In this way,  the weighted mean and covariance of the discrete distribution $\{ \mathcal{X}_i, i=0, 1, \dotsb, 2L \}$, 
\begin{equation}  \label{weighted Pxx}
\begin{split}
&\hat{\mathcal{X}} = \sum\limits_{i=0}^{2L} W_i \mathcal{X}_i \, , \\
& \hat{\mathbf{P}}_{\mathcal{XX}} = \sum\limits_{i=0}^{2L} W_i \left(\mathcal{X}_i-\hat{\mathcal{X}} \right) \left(\mathcal{X}_i-\hat{\mathcal{X}} \right)^T \, , \\
\end{split}
\end{equation}
are exactly the same as $\hat{\mathbf{x}}$ and $\hat{\mathbf{P}}_{xx}$. If $\mathbf{x}$ follows a Gaussian distribution, then $\lambda$ can be chosen as $\lambda =3 - L$ in order to let the first four weighted moments of the discrete distribution $\{ \mathcal{X}_i, i=0, 1, \dotsb, 2L \}$  match those of the distribution of  $\mathbf{x}$ \cite{Julier2000,Julier2004}. 

Because of the symmetry in the sigma points, the rank of the matrix $\hat{\mathbf{P}}_{\mathcal{XX}}$ is $L$. To avoid rank deficiency in the sample covariance matrix, it is suggested that the number of the sigma points be larger than twice the dimension of the vector $\mathbf{x}$ \cite{Julier2000,Julier2004}, or equivalently, $L \ge m$.

After the transformation, the weighted sample mean and covariance of the transformed sigma points are calculated according to
\begin{subequations}
\begin{align}
\label{ut mean} &\hat{\mathbf{y}} = \sum\limits_{i=0}^{2L} W_i \, \mathbf{f} \left( \mathcal{X}_{i} \right) \, ,  \\
 \label{ut cov} &\hat{\mathbf{P}}_{yy} =  \sum\limits_{i=0}^{2L} W_i \, \left( \mathbf{f} \left( \mathcal{X}_{i} \right) - \hat{\mathbf{y}} \right) \left(  \mathbf{f} \left( \mathcal{X}_{i} \right) - \hat{\mathbf{y}} \right ) ^T 
 + \beta \left( \mathbf{f} \left( \mathcal{X}_{0} \right) - \hat{\mathbf{y}} \right) \left(  \mathbf{f} \left( \mathcal{X}_{0} \right) - \hat{\mathbf{y}} \right ) ^T \, ,
\end{align}
\end{subequations}
where the second term on the rhs of Eq. (\ref{ut cov}) is introduced to reduce the approximation error. In the case that $\mathbf{x}$ follows a Gaussian distribution, the choice of $\beta=2$ is shown to be optimal \cite{Julier2004}.

To take into account the effect of the model error in a more general
situation, wherein the system model is described by $\mathbf{y} =
\mathbf{f} \left(\mathbf{x}, \mathbf{u} \right)$ (such that the model
error $\mathbf{u}$ is possibly not additive, but is assumed to follow a Gaussian process 
with mean zero and covariance $\mathbf{Q}$), it is customary to
adopt the joint state $\mathbf{z} = [\mathbf{x}^T, \mathbf{u}^T ]^T$
and the system model is changed to $\mathbf{y} = \mathbf{f}
\left(\mathbf{z} \right)$. In this case, the sigma points can be generated according to
\begin{equation} \label{joint state ut}
\begin{split}
& \mathcal{Z}_0 = \left[\hat{\mathbf{x}}^T, \mathbf{0}^T \right]^T \\
& \mathcal{Z}_i = \mathcal{Z}_0 + \left( \sqrt{(L+\lambda) \hat{\mathbf{P}}_{zz}} \right)_i, \, i=1, 2, \dotsb, L ,\\
& \mathcal{Z}_i = \mathcal{Z}_0 - \left( \sqrt{(L+\lambda) \hat{\mathbf{P}}_{zz}} \right)_{i-L}, \, i=L+1, L+2, \dotsb, 2L ,\\
\end{split}
\end{equation}
where $\mathbf{0}$ means the zero vector, and
$\hat{\mathbf{P}}_{zz}$ is the sample covariance matrix of
$\mathbf{z}$ such that
\begin{equation}
\hat{\mathbf{P}}_{zz} = \begin{pmatrix} \hat{\mathbf{P}}_{xx} & \mathbf{0} \\ \mathbf{0} & \mathbf{Q} \end{pmatrix}
\end{equation}

We will leave the awkward analysis of the estimation accuracies of
both the ordinary EnKF and the unscented transform to the Appendix, where
the general form $\mathbf{y} = \mathbf{f} \left(\mathbf{z} \right)$
described in terms of joint state is considered. We will show that, in
the estimation scheme of the ordinary EnKF, the random samples will
generally introduce spurious modes in the transformed distribution
even if the set of sample points has the correct mean and
covariance \cite{Julier2000,Julier2004}, while for the unscented transform,
by carefully choosing the samples (i.e. the sigma points), the
effect of sample error can be reduced. This fact leads to our
argument that incorporating the unscented transform  may benefit the performance of the EnKF. In the next two sections we will firstly introduce the modification scheme which incorporates the unscented transform for large-scale problems, 
and then proceed to conduct some numerical experiments to examine its performance.


\section{The ensemble unscented Kalman filter}\label{sec:EnUKF}

For large-scale problems, it is not practical to fulfil the requirement that the number of the sigma points should be larger than twice the degrees-of-freedom of the system model \cite{Han-evaluation}. Therefore we will introduce some modifications to make the unscented transform applicable for those problems. We will call the EnKF equipped with the modifications the ensemble unscented Kalman filter (EnUKF). For consistency, we again take Eqs. (\ref{system model}) and (\ref{observer}) as the $m$-dimensional system model and  the $p$-dimensional observer respectively. Moreover, we also assume that there is a set of sigma points  $\{\mathcal{X}_{k-1,i}^a, i=0, 1, \dotsb, 2l_{k-1} \}$ available at the $(k-1)$-th step, which is associated with the weights $W_i$'s given by Eq. (\ref{ut weights}). If there is only an ``ordinary'' ensemble of the analysis available, one may first compute the sample covariance, and then conduct the truncated singular value decomposition (TSVD) \cite{Hansen-truncated} to construct a set of sigma points, as to be explained later.   

With the zero mean of the dynamical noise, at the $k$-th cycle, there exists a set of the predictions of the propagated sigma points $\{\mathcal{X}_{k,i}^b:  \mathcal{X}_{k,i}^b = \mathcal{M}_{k-1,k} \left( \mathcal{X}_{k-1,i}^a \right) , i=0, 1, \dotsb, 2l_{k-1} \}$, which is associated with a set of weights $\left\{ W_{k-1,0}, \dotsb, W_{k-1,2l_{k-1}}\right\}$ obtained from the previous cycle (to be discussed later). The weighted sample mean and covariance are given by
\begin{subequations}
\begin{align}
\label{enukf mean} \hat{\mathbf{x}}_{k}^b = & \sum\limits_{i=0}^{2l_{k-1}} W_{k-1,i} \, \mathcal{X}_{k,i}^b \, ,  \\
 \label{enukf cov} \hat{\mathbf{P}}_{k}^b =  & \sum\limits_{i=0}^{2l_{k-1}} W_{k-1,i} \, \left(\mathcal{X}_{k,i}^b - \hat{\mathbf{x}}_{k}^b \right) \left(  \mathcal{X}_{k,i}^b - \hat{\mathbf{x}}_{k}^b \right ) ^T + \\
\nonumber &\beta \left(\mathcal{X}_{k,0}^b - \hat{\mathbf{x}}_{k}^b \right) \left(  \mathcal{X}_{k,0}^b - \hat{\mathbf{x}}_{k}^b \right ) ^T + \mathbf{Q}_k \, .
\end{align}
\end{subequations}

Moreover, the above evaluation scheme is also applied to the projection of the background ensemble such that
\begin{equation} \label{enukf background covariances}
\begin{split}
\hat{\mathbf{y}}_k^b = & \sum\limits_{i=0}^{2l_{k-1}} W_{k-1,i} \mathcal{H}_k \left ( \mathcal{X}_{k,i}^b \right) ,  \\
\hat{\mathbf{P}}_{xh}^k = & \sum\limits_{i=0}^{2l_{k-1}} W_{k-1,i} \, \left( \mathcal{X}_{k,i}^b - \hat{\mathbf{x}}_{k}^b \right) \left( \mathcal{H}_k \left ( \mathcal{X}_{k,i}^b \right )- \hat{\mathbf{y}}_k^b \right ) ^T \\
&+ \beta \left(\mathcal{X}_{k,0}^b - \hat{\mathbf{x}}_{k}^b \right) \left( \mathcal{H}_k \left ( \mathcal{X}_{k,0}^b \right )- \hat{\mathbf{y}}_k^b \right ) ^T , \\
\hat{\mathbf{P}}_{hh}^k =  & \sum\limits_{i=0}^{2l_{k-1}} W_{k-1,i} \, \left( \mathcal{H}_k \left ( \mathcal{X}_{k,i}^b \right )- \hat{\mathbf{y}}_k^b \right ) \left( \mathcal{H}_k \left ( \mathcal{X}_{k,i}^b \right )- \hat{\mathbf{y}}_k^b\right ) ^T \\
&+ \beta  \left( \mathcal{H}_k \left ( \mathcal{X}_{k,0}^b \right )- \hat{\mathbf{y}}_k^b\right ) \left( \mathcal{H}_k \left ( \mathcal{X}_{k,0}^b \right )- \hat{\mathbf{y}}_k^b\right ) ^T. \\
\end{split}
\end{equation}

For numerical reason, it is customary to re-write the above error covariances in terms of some square root matrices. To this end, we introduce two square roots, $\mathbf{S}^{x}_k$ and $\mathbf{S}^{h}_k$, which are defined by 
\begin{subequations}
\begin{align}
\label{sut SRX} \mathbf{S}^{x}_k = & \left[ \sqrt{W_{k-1,0}^{\beta}} \left( \mathcal{X}_{k,0}^b - \hat{\mathbf{x}}_{k}^b \right), \sqrt{W_{k-1,1}} \left( \mathcal{X}_{k,1}^b - \hat{\mathbf{x}}_{k}^b \right), \right. \\
\nonumber &\quad \left. \dotsb, \sqrt{W_{k-1, 2l_{k-1}}} \left( \mathcal{X}_{k,2l_{k-1}}^b - \hat{\mathbf{x}}_{k}^b \right) \right], \\
\label{sut SRH} \mathbf{S}^{h}_k = & \left[ \sqrt{W_{k-1,0}^{\beta}} \left( \mathcal{H}_k \left ( \mathcal{X}_{k,0}^b \right )- \hat{\mathbf{y}}_{k} \right), \sqrt{W_{k-1,1}} \left( \mathcal{H}_k \left (\mathcal{X}_{k,1}^b \right )- \hat{\mathbf{y}}_{k} \right), \right. \\
\nonumber &\quad \left. \dotsb, \sqrt{W_{k-1, 2l_{k-1}}} \left( \mathcal{H}_k \left ( \mathcal{X}_{k,2l_{k-1}}^b \right )- \hat{\mathbf{y}}_{k} \right) \right],
\end{align}
\end{subequations}
where $W_{k-1,0}^{\beta} = W_{k-1,0}+\beta$. Then the covariances can be re-written as
\begin{subequations}
\begin{align}
\label{sut SQ cov} & \hat{\mathbf{P}}_k^b = \mathbf{S}^{x}_k \left( \mathbf{S}^{x}_k \right)^T + \mathbf{Q}_k,\\
\label{sut SQ cross cov} & \hat{\mathbf{P}}^{k}_{xh} =  \mathbf{S}^{x}_k \left( \mathbf{S}^{h}_k \right)^T, \\
\label{sut SQ projection cov} & \hat{\mathbf{P}}^{k}_{hh} =   \mathbf{S}^{h}_k \left( \mathbf{S}^{h}_k \right)^T, 
\end{align}
\end{subequations}

Again, the Kalman gain follows Eq. (\ref{Kalman gain}), i.e.,
\begin{equation} \tag{$\ref{Kalman gain}$}
\mathbf{K}_k = \hat{\mathbf{P}}_{xh}^k \left( \hat{\mathbf{P}}_{hh}^k + \mathbf{R}_k \right)^{-1}.
\end{equation}

With the above information, the mean and covariance of the analysis can be computed according to
\begin{subequations}
\begin{align}
\label{enukf: analysis mean} & \hat{\mathbf{x}}_k^{a} =  \hat{\mathbf{x}}_k^{b} + \mathbf{K}_k  \left( \mathbf{y}_k - \mathcal{H}_k \left( \hat{\mathbf{x}}_k^{b} \right) \right),\\
\label{enukf: analysis cov} & \hat{\mathbf{P}}_k^a = \hat{\mathbf{P}}_k^b -  \mathbf{K}_k \left( \hat{\mathbf{P}}_{xh}^k \right)^T .
\end{align}
\end{subequations}

Apart from obtaining the updated sample mean and covariance, we also aim to generate a set of sigma points as the analysis ensemble, which will then be propagated to the next assimilation cycle. For this purpose, one may consider using an existing EnKF scheme, for example, the ETKF. However, in order to avoid doubling the ensemble size at each assimilation cycle, some sigma points have to be discarded. To do this, the sample mean can be preserved by maintaining the symmetry about $\hat{\mathbf{x}}_k^{a}$ among the remaining sigma points, while the corresponding sample covariance, denoted by $\tilde{\mathbf{P}}_k^a$, can only be an approximation to $\hat{\mathbf{P}}_k^a$. This may appear to be a complicated problem for the existing EnKFs to design a selection criterion, because the perturbations produced by them have no indications of the relative importance for covariance approximation. 

To tackle the above problem, the truncated singular value decomposition (TSVD) \cite{Hansen-truncated} is adopted in this work, which is similar to the idea of using singular vectors to produce initial perturbations for ensemble forecasting (\cite{Ehrendorfer1997}, \cite[ch 6]{Kalnay-atmospheric}, \cite{Turner-ensemble}, \cite{Uzunoglu-adaptive}, and the references therein). 
Specifically, to produce the sigma points, a singular value decomposition (SVD) is firstly conducted on $\hat{\mathbf{P}}_k^a$. Suppose that  $\hat{\mathbf{P}}_k^a$ can be expressed as
\begin{equation}
\hat{\mathbf{P}}_k^a = \mathbf{E}_k \mathbf{D}_K \left(\mathbf{E}_k\right)^T ,
\end{equation}
where $\mathbf{D}_K = \text{diag} (\sigma_{k,1}^2, \dotsb, \sigma_{k,m}^2)$ is a diagonal matrix consisting of the eigenvalues $\sigma_{k,i}^2$'s of $\hat{\mathbf{P}}_k^a$, which are sorted in descending order, i.e., $\sigma_{k,i}^2 \ge \sigma_{k,j}^2 \ge 0$ for $i>j$ , and $\mathbf{E}_K = \left[\mathbf{e}_{k,1}, \dotsb,  \mathbf{e}_{k,m} \right] $  is the matrix consisting of the corresponding eigenvectors $\mathbf{e}_{k,i}$'s. Next, a set of perturbations, generated in terms of the first $l_k$ values of $\sigma_{k,i} \mathbf{e}_{k,i}$, are added to the sample mean $\hat{\mathbf{x}}_k^{a}$ to form $l_k$ sigma points. Another $l_k$ symmetric sigma points can be produced by subtracting the perturbations from the sample mean. Overall, in analogy to Eq. (\ref{sigma points}), the above procedure can be summarized as follows
\begin{equation} \label{enukf: sigma point generation}
\begin{split}
& \mathcal{X}_{k,0}^a = \hat{\mathbf{x}}_k^{a},\\
& \mathcal{X}_{k,i}^a =  \hat{\mathbf{x}}^{a}_k + (l_k+\lambda)^{1/2} \sigma_{k,i} \mathbf{e}_{k,i}, \, i=1, \dotsb, l_k ,\\
& \mathcal{X}_{k,i}^a =  \hat{\mathbf{x}}^{a}_k -  (l_k+\lambda)^{1/2} \sigma_{k,i-l_k} \mathbf{e}_{k,i-l_k}, \, i=l_k+1, \dotsb, 2l_k ,\\
\end{split}
\end{equation}
where $\lambda$ is the adjustable scaling parameter. Using the generated sigma points as the analysis ensemble, the EnUKF is clearly an unbiased ensemble filter \cite{Livings-unbiased}. 

It is worthy to note that, Eq. (\ref{enukf: sigma point generation}) does not require the full spectra of the eigenvalues and eigenvectors. Therefore, to reduce the computational cost in large scale problems, some fast SVD algorithms, e.g.,  the Lanczos or block Lanczos algorithm \cite[ch 9]{Golub-matrix}, can be adopted to compute the first $l_k$ pairs of eigenvalues and eigenvectors (for example, see \cite{Treebushny-construction}).

For convenience, we will hereafter call $l_k$ the truncation number. The choice of $l_k$ is important to the performance of the EnUKF, because it not only controls the number of sigma points to be produced, but also determines the quality of matrix approximation. Indeed, via SVD the matrices $\hat{\mathbf{P}}_k^a$ and $\tilde{\mathbf{P}}_k^a$ can be expressed as
\begin{equation}
\begin{split}
&\hat{\mathbf{P}}_k^a = \sum\limits_{i=1}^{m} \sigma_{k,i}^2 \mathbf{e}_{k,i} \left( \mathbf{e}_{k,i} \right)^T \, ,\\
&\tilde{\mathbf{P}}_k^a = \sum\limits_{i=1}^{l_k} \sigma_{k,i}^2 \mathbf{e}_{k,i} \left( \mathbf{e}_{k,i} \right)^T \, ,\\
\end{split}
\end{equation}
respectively. It is clear that, if $l_k$ is too small, some important structures of  $\hat{\mathbf{P}}_k^a$, in terms of $\sigma_{k,i}^2 \mathbf{e}_{k,i} \left( \mathbf{e}_{k,i} \right)^T$ for $i > l_k$, will be lost. However, as the computational cost is also a concern, it is not desirable for $l_k$ to get too large. Moreover, in many situations, if $l_k$ is large enough, $\sigma_{k,l_k}^2$ may be already very small compared to the leading eigenvalues. Thus the improvement obtained by increasing $l_k$ becomes negligible. In this sense, one may choose a modest value of $l_k$ to achieve a tradeoff between accuracy and efficiency.  
 
In our implementation, we let $l_k$ be an integer such that
\begin{equation} \label{enukf:threshold}
\begin{split}
& \sigma_{k,i}^2 > \text{trace} \left( \hat{\mathbf{P}}_k^a \right) / h_k \, , i=1, \dotsb, l_k \, ,\\
& \sigma_{k,i}^2 \le \text{trace} \left( \hat{\mathbf{P}}_k^a \right) / h_k \, ,  i>l_k +1 \, ,
\end{split}
\end{equation}
where $h_k$ is the threshold at the $k$-th cycle (we will discuss how to choose $h_k$ in section \ref{sec:ukf config}). This is equivalent to saying that we construct the sigma points based on the eigenvectors such that their corresponding eigenvalues are larger than a specified tolerance. Moreover, to prevent $l_k$ getting too large or too small, we also specify a lower bound $l_l$ and an upper bound $l_u$ and adjust the threshold $h_k$ so that $l_l \le l_k \le l_u$. 

Under the assumption of Gaussian error distribution, it can be verified that the perturbations of the sigma points, in terms of $(l_k+\lambda)^{1/2} \sigma_{k,i} \mathbf{e}_{k,i}$ for $i=1,\dotsb, l_k$, are equally likely in the sense that their values of the probability density function (PDF) 
\begin{equation}
p(\delta \mathbf{x}) = \left( 2 \pi \right)^{m/2} (\det \hat{\mathbf{P}}_k^a)^{-1/2} \text{exp} \left\{ -\dfrac{1}{2} \left( \delta \mathbf{x}\right)^T \left( \hat{\mathbf{P}}_k^a \right)^{-1} \left( \delta \mathbf{x}\right) \right\} 
\end{equation}
are the same (also see the discussions in \cite{Wang-which}), where $\det \bullet$ means the determinant of a matrix. Therefore it is natural to assign an identical weight to all the perturbations. Consequently, in the spirit of Eq. (\ref{ut weights}), a set of weights can be constructed as follows
\begin{equation} \label{enukf:weights}
\begin{split}
&W_{k,0} = \frac{\lambda}{l_k+\lambda},\\
&W_{k,i} = \frac{1}{2\left(l_k+\lambda  \right)}, \, i=1, \dotsb, 2l_k. \\
\end{split}
\end{equation}
Finally, all the sigma points in Eq. (\ref{enukf: sigma point generation}), which are associated with a set of weights given in Eq. (\ref{enukf:weights}), are propagated forward to the next assimilation cycle.

We adopt the time averaged relative rms error (relative rmse for short) to measure the performance of the EnUKF, which is defined as
\begin{equation}\label{Eq:reltiave rmse}
e_r=\frac{1}{k_{max}}\sum\limits_{k=1}^{k_{max}} \lVert\hat{\mathbf{x}}_k^{a}-{\mathbf{x}}_k^{tr}\rVert_2/ \lVert{\mathbf{x}}_k^{tr}\rVert_2,
\end{equation}
where $k_{max}$ is the maximum assimilation cycle, $\mathbf{x}_k^{tr}$ denotes the truth (the state of a control run) at the $k$-th cycle, and $\lVert \bullet \rVert_2$ means the $L_2$ norm. 

Moreover,  we also use the time averaged rms ratio to examine the similarity of the truth to the sigma points, which also qualitatively reflects the performance in estimating the error covariance, e.g., overestimation or underestimation (cf. \cite{Anderson-ensemble, Whitaker-ensemble} and the references therein). As an estimation, the time averaged rms ratio, denoted by $R$, is computed by \cite{Anderson-ensemble, Whitaker-ensemble}
\begin{equation}
R =\frac{1}{k_{max}} \sum\limits_{k=1}^{k_{max}}  \left(2l_k+1\right) \lVert\hat{\mathbf{x}}_k^{a}-{\mathbf{x}}_k^{tr}\rVert_2 / \sum\limits_{i=0}^{2l_k} \lVert\mathcal{X}_{k,i}^{a}-{\mathbf{x}}_k^{tr}\rVert_2  \, ,
\end{equation}
while the expectation of the rms ratio is \cite{Anderson-ensemble, Whitaker-ensemble}
\[
R_e = \sqrt{(l_{eff}+1)/(2l_{eff}+1)} \, ,
\]
where $l_{eff}$ is the ``effective'' truncation number over the whole assimilation window. Hence, if the truth is statistically indistinguishable from the sigma points, the values of $R$ and $R_e$ shall be very close. Note that $R_e \approx 0.71$ for any large $l_{eff}$, so for simplicity we let $l_{eff}$ equal the average of the truncation number $\bar{l}$, i.e., $l_{eff} = \bar{l} = \sum_{i=1}^{k_{max}} l_k / k_{max}$. $R>R_e$ means that the covariance of the sigma points underestimates the error of the state estimation, while $R<R_e$ implies the opposite, i.e., overestimation of the error of of the state estimation \cite{Murphy-impact, Whitaker-ensemble}.

\section{Numerical experiments with a 40-dimensional system} \label{sec:results}

This section is dedicated to examining the performance of the EnUKF through the numerical simulations, and studying the effects of the parameters on the performance of the EnUKF. To this end, we choose the $m$-dimensional system model due to Lorenz and Emanuel \cite{Lorenz-predictability,Lorenz-optimal} (LE98 model hereafter) as the testbed. The LE98 model is a simplified system proposed to model atmospheric dynamics, which ``shares certain properties with many atmospheric models'' \cite{Lorenz-optimal}. We consider the perfect model scenario, wherein the governing equations are described as follows
\begin{equation} \label{LE98}
\frac{dx_i}{dt} = \left( x_{i+1} - x_{i-2} \right) x_{i-1} - x_i +F, \, i=1, \dotsb, m. 
\end{equation}
The quadratic terms simulate the advection, the linear term represents the internal dissipation, while the constant $F$ acts as the external forcing (\cite{Lorenz-predictability}). The variables $x_i$'s are defined cyclically such that $x_{-1}=x_{m-1}$, $x_{0}=x_{m}$, and $x_{m+1}=x_{1}$. 

We choose the observer $\mathcal{H}_k$ to be a time-invariant identity operator. Specifically, given a system state $\mathbf{x}_k=[x_{k,1},\dotsb,x_{k,m}]^T$ at the $k$-th assimilation cycle, the observations are obtained according to
\begin{equation} \label{sim:observer}
\mathbf{y}_k = \mathcal{H}_k (\mathbf{x}_k) + \mathbf{v}_k = \mathbf{x}_k + \mathbf{v}_k \, ,
\end{equation} 
where $\mathbf{v}_k$ follows the $m$-dimensional Gaussian distribution $N(\mathbf{0}, \mathbf{R}_k)$ with the covariance matrix $\mathbf{R}_k$ being the $m \times m$ identity matrix $\mathbf{I}_m$. 

Note that in Eq.~(\ref{Eq:reltiave rmse}) the measure $e_r$ can also be interpreted as the time-averaged noise level of the trajectory $\{ \hat{\mathbf{x}}_k^{a} \}_{k=1}^{k_{max}}$ (with respect to the truth). From this point of view, we can define the divergence of a filter in a restrictive sense: Suppose that the relative rmse of the observations is $e_r^{obv}$, which, with the identity observation operator $\mathcal{H}_k$, is defined as $\lVert\mathbf{y}_k-{\mathbf{x}}_k^{tr}\rVert_2/ \lVert{\mathbf{x}}_k^{tr}\rVert_2 = \lVert\mathbf{v}_k \rVert_2/ \lVert{\mathbf{x}}_k^{tr}\rVert_2$ . If $e_r>e_r^{obv}$, then we say the filter is divergent because in such circumstances, the trajectory $\{ \hat{\mathbf{x}}_k^{a} \}_{k=1}^{k_{max}}$ obtained by the filter, on average, is more noisy than the observations, which implies that it might not make any sense to use the filter for assimilation. 

In our experiments, we set $m=40$ and $F=8$ and integrate the system through the fourth-order Runge-Kutta method. We choose the length of the integration window to be $100$ dimensionless units, and the integration time step to be $0.05$ units (corresponding to about a 6-h interval in reality, see \cite{Lorenz-optimal}), thus there are $2000$ assimilation cycles overall. 

For the LE98 model, in the numerical experiments (not reported there) we found that, the EnSRF, implemented following either work of \cite{Anderson-ensemble,Bishop-adaptive,Whitaker-ensemble}, outperforms the stochastic EnKF, which is consistent with the result reported in \cite{Whitaker-ensemble}, while the performances of the EnSRFs are very close. Therefore, here we choose to compare the EnUKF with the EnSRF only. Moreover, for an ordinary EnSRF, we will introduce the spherical simplex centering scheme to its analysis ensembles. The reason to do this is two-fold. Firstly,  with the spherical simplex centering scheme, the produced analysis perturbations are equally likely in probability under the assumption of Gaussian error distribution \cite{Wang-which}. Secondly, with the centering scheme, an EnSRF can be shown to be an unbiased ensemble filter, which avoids an error that systematically underestimates the analysis covariance (see \cite{Livings-unbiased} for the details). In this work, we single out the ETKF \cite{Bishop-adaptive} for our experiments, and equip it with the spherical simplex centering scheme following the work \cite{Wang-which}, which will be further discussed below. 

\subsection{Issues in implementing the EnUKF and the ETKF}\label{sec:ukf config}

When implementing the EnUKF, an issue worth of special attention is the positive semi-definiteness of the covariance matrices.  Normally, we require $\l_k + \lambda >0$ so that in Eq. (\ref{enukf:weights}), the weights $W_{k,i}$'s are positive for $i>0$. Nevertheless, the weight $W_{k,0}$ can be negative if $\lambda<0$. If it is so, when computing the background covariances according to Eqs. (\ref{enukf cov}) and (\ref{enukf background covariances}), the positive semi-definiteness may not be guaranteed. However, one may note that, in Eqs. (\ref{enukf cov}) and (\ref{enukf background covariances}), the effective weight of $ \mathcal{X}_{k+1,0}^b$ is actually $W_{k,0}+\beta$ ($\beta \ge 0$). So in order to guarantee the positive semi-definiteness, we shall choose the parameters $\lambda$ and $\beta$ properly to satisfy that $W_{k,0}+\beta \ge 0$ and $l_k +\lambda >0$. Given $W_{k,0} = \lambda / (l_k + \lambda)$, it means that $\lambda \ge - \beta l_k / (1+\beta)$. Since $l_k$ is bounded within $\left[l_l, l_u \right]$, by letting $\lambda \ge - \beta l_l / (1+\beta)$, one can guarantee the positive semi-definiteness. 

The threshold $h_k$ in Eq. (\ref{enukf:threshold}) is chosen in the following way. At the beginning we specify a threshold $h_1$. If $h_1$ is a proper value such that $l_1$ satisfies $l_l \le l_1 \le l_u$, then we keep $h_1$ and at the next cycle we start with $h_2=h_1$. If $h_1$ is too small such that $l_1<l_l$, then we replace $h_1$ by $1.1 h_1 + 200$. We continue the replacement until $l_1$ falls into the specified range, or the number of the replacement operations is up to $30$ (in this case we simply put $l_1=l_l$, regardless of what $h_1$ is). Similarly, if $h_1$ is too large such that $l_1>l_u$, then we replace $h_1$ by $h_1/1.1 - 200$, we continue the replacement until $l_1$ falls in the specified range, or the number of the operations is up to $30$ (in this case we simply put $l_1=l_u$). After the adjustment, at the next cycle we start with $h_2=h_1$ and adjust it (if necessary) to let $l_2$ fall into the specified range, and so on. In this way, one can obtain the threshold at each cycle.

To apply the spherical simplex centering scheme to the ETKF, we follow the algorithm proposed in \cite{Julier2004} to construct a centering matrix $\mathbf{U}$, where $\mathbf{U}$ follows Eq. (C15) in \cite{Wang-which}. Compared to the alternative centering matrix provided in \cite{Wang-which}, we favor the one described in Eq. (C15) because it is time-invariant. Moreover, its construction is independent of the concrete system in assimilation, and does not involve the operation of matrix inversion and the observation operator. 

In our experiments, we process the observations simultaneously in both the EnUKF and the ETKF. In order to improve the performances of the filters, we also consider two additional techniques. One is the method of covariance inflation, which is based on the observation that the covariance of the analysis error will be systematically underestimated in the EnKF \cite{Whitaker-ensemble}. Therefore, it can be beneficial to increase either the background error covariance, or the analysis error covariance \cite{Anderson-Monte,Ott-local,Whitaker-ensemble}. In this work, we follow the method used in \cite{Anderson-Monte,Whitaker-ensemble} and choose to multiply the perturbations to the sample mean $\mathbf{x}_k^a$ of the analysis by a constant $1+\delta$, which is equivalent to increasing the analysis error covariance by a factor $(1+\delta)^2$. Note that, if one chooses to process the observations in a serial way (e.g. \cite{Whitaker-ensemble}) with a covariance factor $\delta_s$, then after processing an $m$-dimensional observation, the corresponding covariance will be increased by a factor of $(1+\delta_s)^{2m}$. The relationship between the inflation factors $\delta$ and $\delta_s$ is thus given by $\delta = (1+\delta_s)^{m}-1$, therefore one will find that the inflation factor adopted in this work is much larger than those used in, for example, \cite{Whitaker-ensemble}. 

\begin{figure*}[htb] 
     \centering
     \subfigure[Relative rmse vs the length scale of covariance filter for the EnUKF]{
        \label{fig:ukf_RRmse_vs_lc}
        \includegraphics[width=\textwidth]{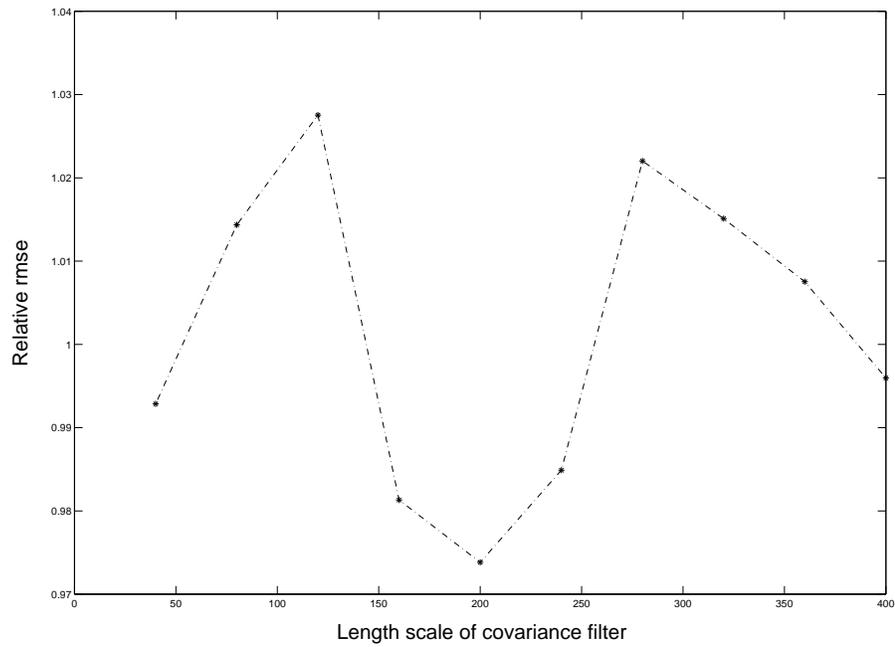}}\qquad
     \subfigure[Relative rmse vs the length scale of covariance filter for the ETKF]{
          \label{fig:srf_RRmse_vs_lc}
          \includegraphics[width=\textwidth]{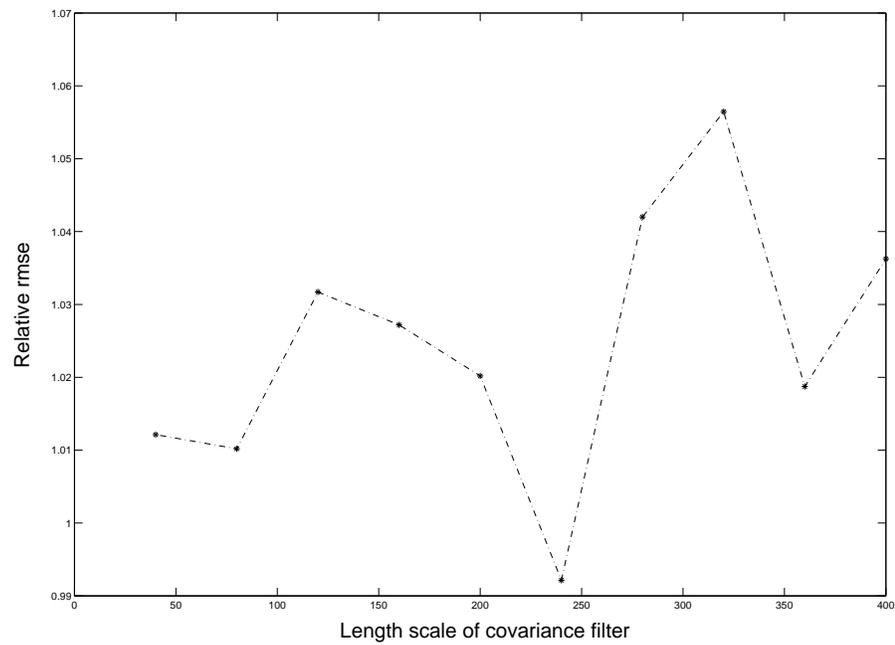}}
     \hspace{.3in}
	
     \caption{ \label{fig:optimal lc} Effects of the length scale of covariance filter on the performances of EnUKF and the ETKF. From
the figures, the optimal length scale of the EnUKF is $200$, while the optimal length scale of the ETKF is $240$.}
\end{figure*}

The other technique is covariance filter \cite{Hamill-distance,Houtekamer-sequential}, which introduces the Schur-product to a covariance matrix in order to reduce the effect of sample errors. We say a length scale of covariance filter is optimal within a certain range if it minimizes the relative rmse among the values in test. As an example, in Fig. \ref{fig:optimal lc} we plot the relative rms errors of the EnUKF (upper panel) and the ETKF (lower panel) vs the length scale. Both filters start with the same initial condition, and takes no covariance inflation (i.e., $\delta=0$). The length scale is varied from $40$ to $400$, with an even increment of $40$ at each step.  The other settings of the filters are as follows: For the EnUKF (upper panel), the initial ensemble size is $4$. $\beta=2$, $\lambda=-2$, the lower bound $l_l=3$, the upper bound $l_u=6$, and the threshold $h_1=1000$; For the ETKF, the ensemble size is $13$. From Fig. \ref{fig:optimal lc}, it is clear that the optimal length scale of the EnUKF is $200$, while the optimal length scale of the ETKF is $240$. For simplicity, we choose $l_c =240$ for both filters in the subsequent simulations.

\subsection{Comparison between the EnUKF and the ETKF}\label{sec:comparison}

For comparison, we randomly select an initial condition $\mathbf{x}_1$, and use it to start a control run. The observations $\mathbf{Y} = \{ \mathbf{y}_k \}_{k=1}^{2000}$ are obtained by adding Gaussian white noise to the states of the control run at each cycle, in accordance with Eq. (\ref{sim:observer}). In subsequent simulations, both the EnUKF and the ETKF will start with the same initial condition $\mathbf{x}_1$, and use the same observations $\mathbf{Y}$ for assimilation. To initialize the filters, at the first assimilation cycle we randomly generate a background ensemble $\mathbf{X}_1^b = \left \{ \mathbf{x}_{1,i}: i=1, \dotsb, n \right \}$. Given $\mathbf{X}_1^b$ and $\mathbf{x}_1$, the ETKF is already able to start running recursively. For the EnUKF, however, at the first cycle there is no propagated sigma points from the previous cycle. So, similarly to the ETKF, one may use the background ensemble $\mathbf{X}_1^b$ to compute the sample mean and covariance of the analysis, and then generate the sigma points accordingly. After propagating the sigma points forward, the EnUKF can start running recursively from the second cycle.

\begin{figure*}[htb] 
     \centering
     \subfigure[Relative rmse of the EnUKF vs the covariance inflation factor $\delta$]{
        \label{fig:ukf_RRmse_VS_InflationFactor_lownEn}
        \includegraphics[width=\textwidth]{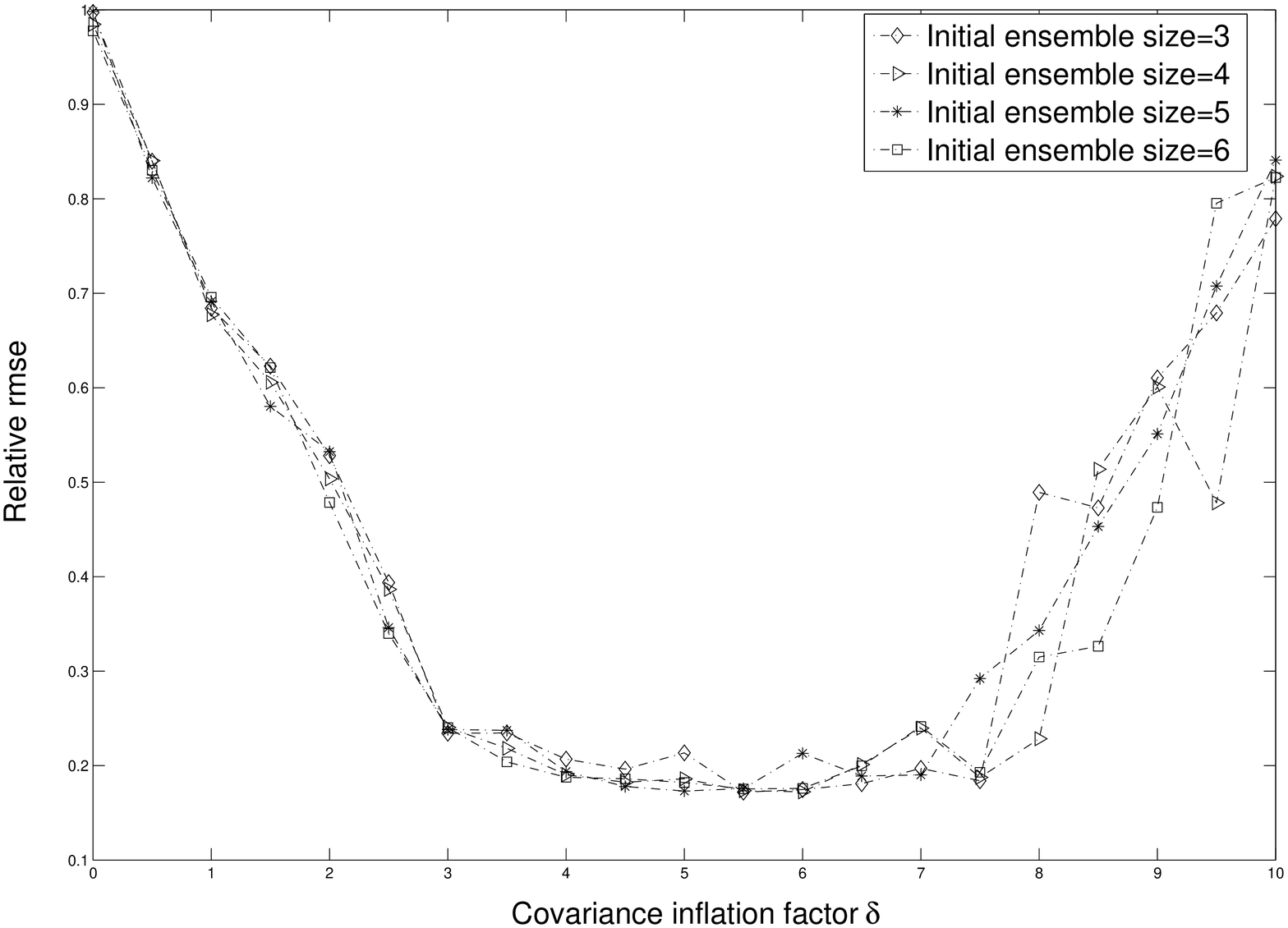}}\qquad
     \subfigure[RMS ratio of the EnUKF vs the covariance inflation factor $\delta$]{
          \label{fig:ukf_rmsRatio_VS_InflationFactor_lownEn}
          \includegraphics[width=\textwidth]{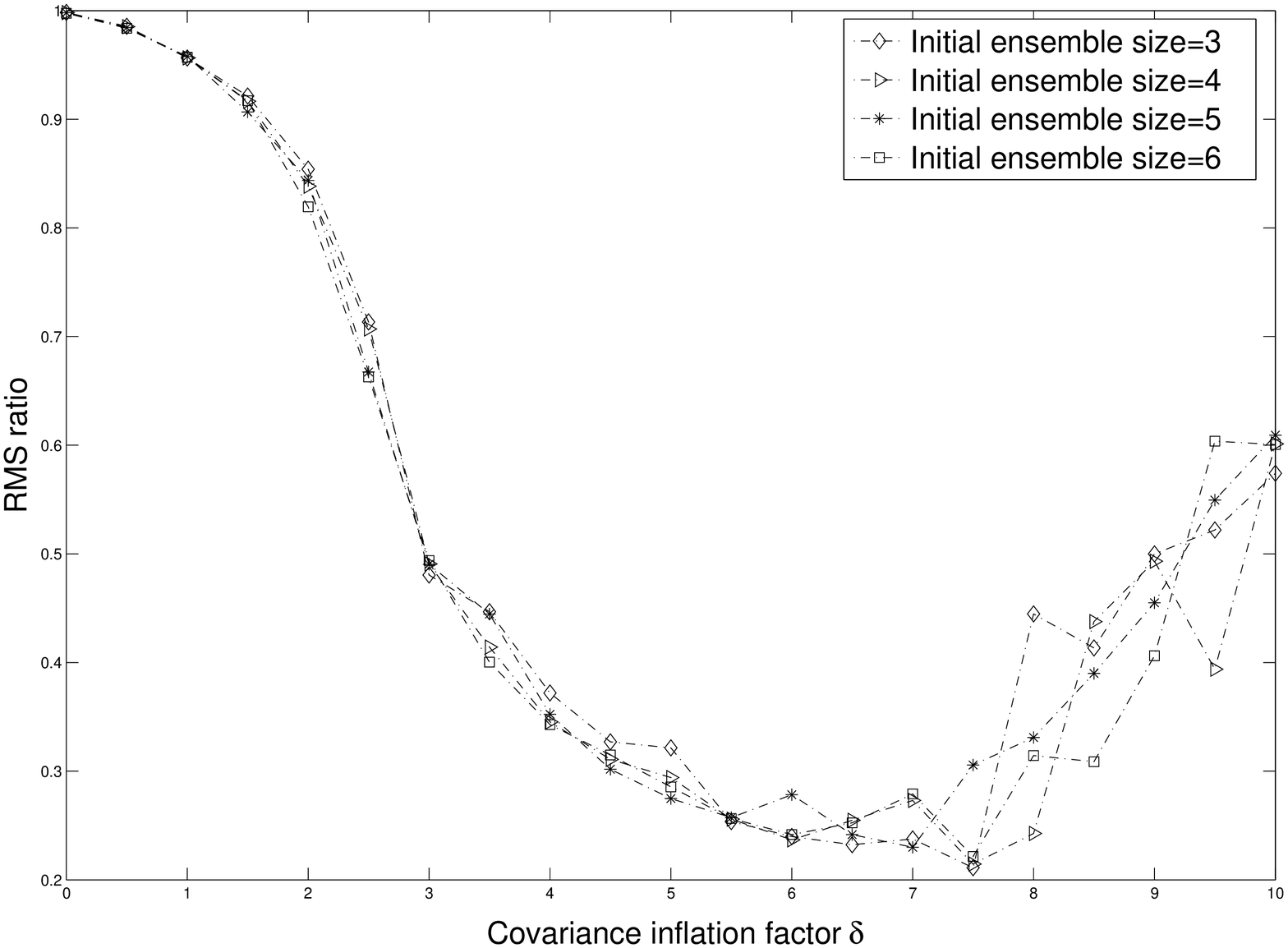}}
     \hspace{.3in}
	
     \caption{ \label{fig:ukf performance} Effects of the covariance inflation factor $\delta$ on the performance of the EnUKF.}
\end{figure*}

For the EnUKF, we let the parameters $\beta=2$, $\lambda=-2$, the threshold $h_1=1000$, the lower bound $l_l=3$, the upper bound $l_u=6$, the length scale of covariance filter $l_c =240$, and the covariance inflation factor $\delta$ vary from $0$ to $10$, with an even increment of $0.5$. We consider the scenarios with different ensemble sizes $n=3,4,5,6$ at the first assimilation cycle in order to explore the effect of initial ensemble size on the performance of the EnUKF. The corresponding relative rms errors and ratios, as functions of the covariance inflation factor, are plotted in Figs. \ref{fig:ukf_RRmse_VS_InflationFactor_lownEn} and \ref{fig:ukf_rmsRatio_VS_InflationFactor_lownEn} respectively

From the above two figures, it can be seen that different initial ensemble sizes $n=3,4,5,6$  leads to similar behaviors of both the relative rmse and rms ratio. Interestingly, a larger initial ensemble size does not necessarily guarantee a smaller rmse error. This can be observed either from Fig. \ref{fig:ukf_RRmse_VS_InflationFactor_lownEn} by fixing the covariance inflation factor $\delta$ at some point, say $\delta=1.5$, or from Table \ref{tab:min rmse} by comparing the minimum relative rms errors. 

Fig. \ref{fig:ukf_RRmse_VS_InflationFactor_lownEn} shows that, as the covariance inflation factor $\delta$ increases from $0$, the relative rmse of the EnUKF tends to decline. However, if $\delta$ gets too large, say $\delta > 6$, then further increments in $\delta$ will instead boost the relative rms errors. An examination on the rms ratio also reveals the same trend, although, as indicated in Fig. \ref{fig:ukf_rmsRatio_VS_InflationFactor_lownEn}, the turning points, now at $\delta =7.5$ for $n=3,4,6$ and $\delta =7$ for $n=5$, are larger than those of the relative rms errors. To make the sigma point indistinguishable from the truth (i.e., rms ratio $\approx 0.71$ ), one needs the inflation factor $\delta \approx 2.5$. However, modestly larger inflation factors, say, $2.5 < \delta <6$, can benefit the performance of the EnUKF in terms of the relative rmse, although they also cause the over-estimations of the error covariances.

\begin{figure*}[htb] 
     \centering 
	\subfigure[Relative rmse of the ETKF with the ensemble size equal to $\text{ceil}(2\bar{l}+1) =13$]{
        \label{fig:enTf_RRmse_vs_delta_meanDoubleL}
        \includegraphics[width=\textwidth]{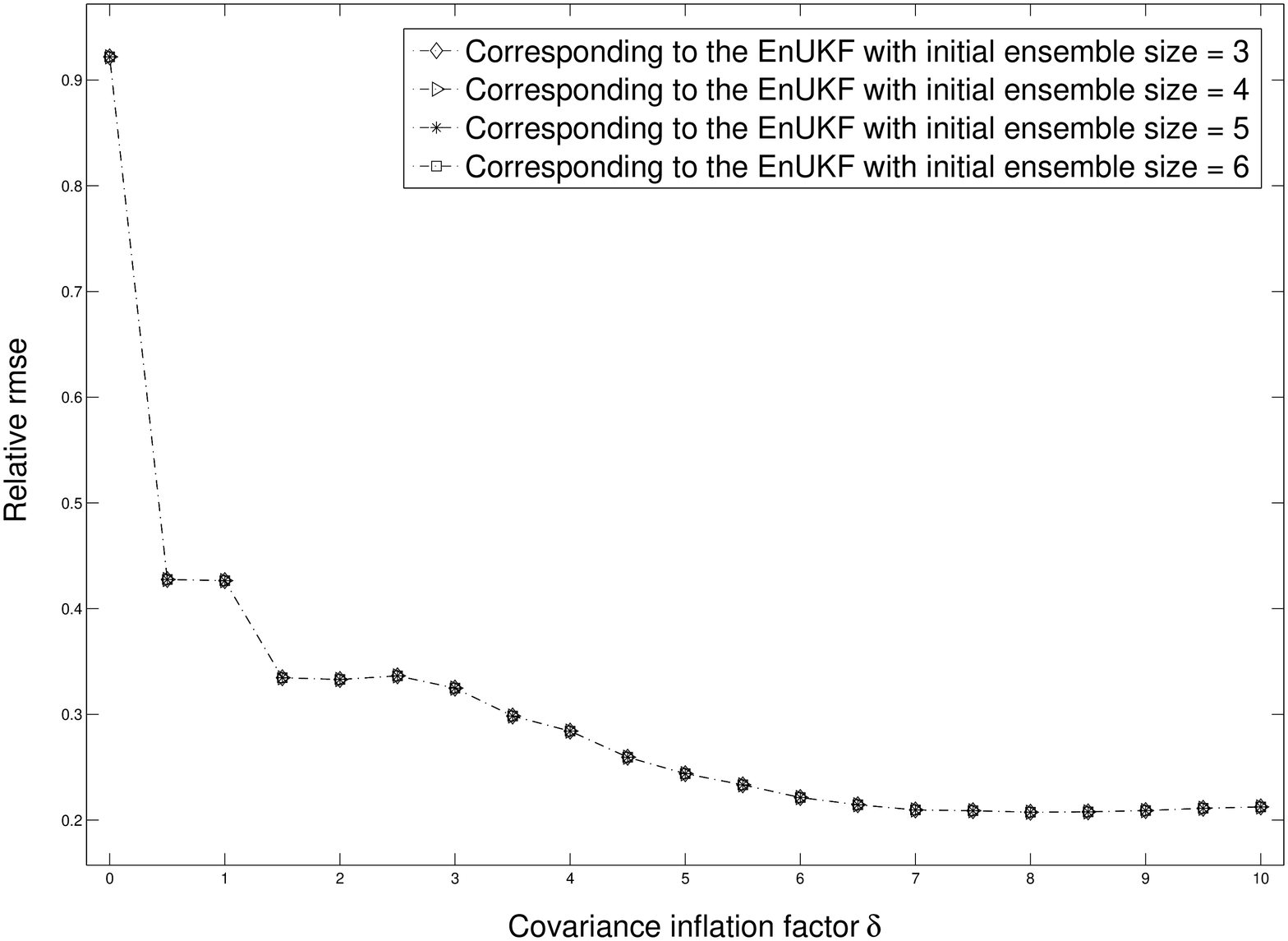}}
     \subfigure[RMS ratio of the ETKF with the ensemble size equal to $\text{ceil}(2\bar{l}+1) =13$]{
          \label{fig:enTf_rmsRatio_vs_delta_meanDoubleL}
          \includegraphics[width=\textwidth]{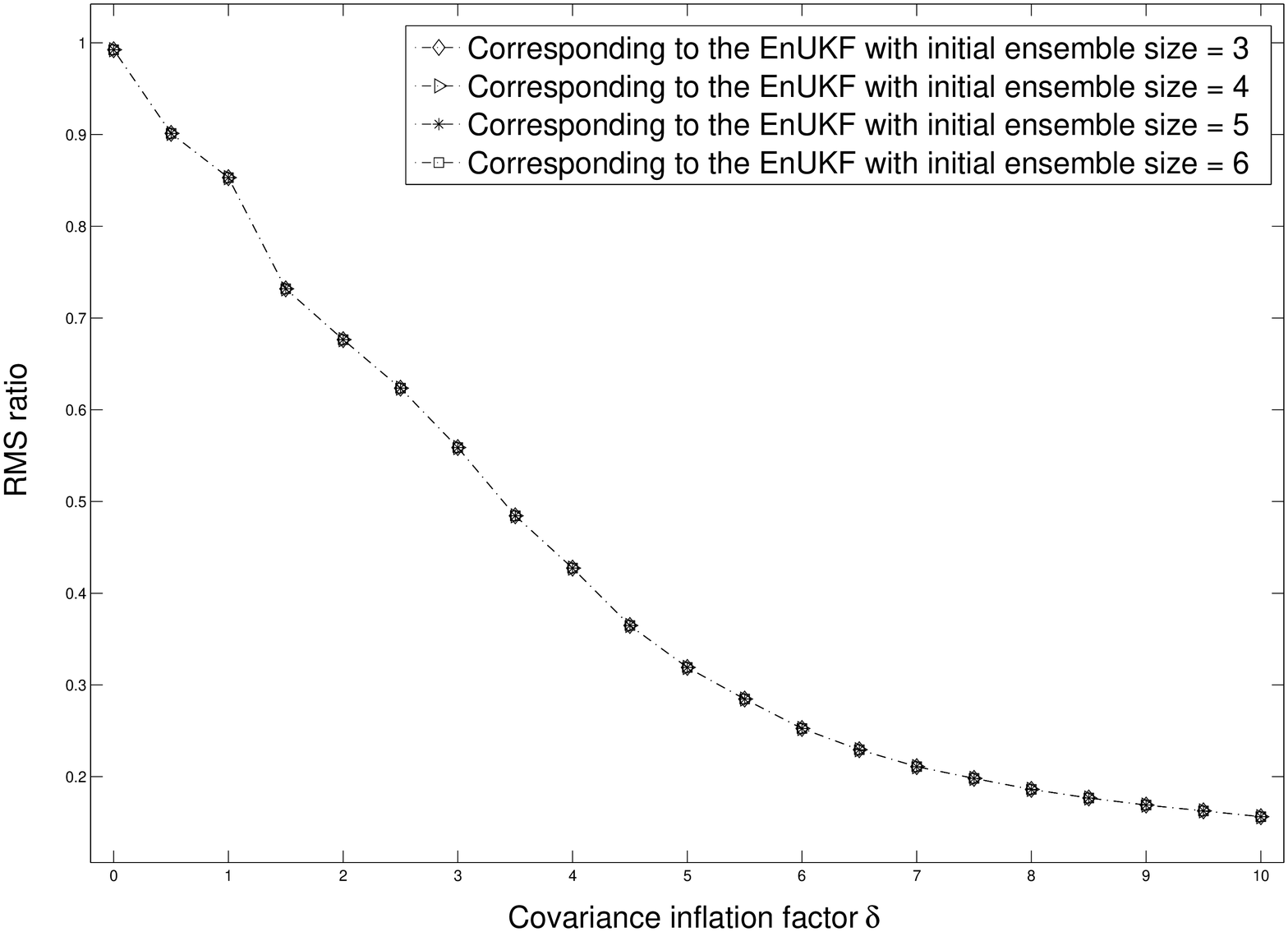}}
     \hspace{.3in}
     \caption{ \label{fig:srf doubleL} Effects of the covariance inflation factor $\delta$ on the performance of the ETKF.}
\end{figure*}

For the ETKF, we let the length scale $l_c$ of covariance filter and the covariance inflation factor $\delta$ be the same as those in the EnUKF. Suppose that in a run of the EnUKF we have the average truncation number $\bar{l}$. Then for comparison, we consider the ETKF with an ensemble size $n=\text{ceil}(2\bar{l}+1)$, where $\text{ceil} (s)$ means the nearest integer that is larger than, or equal to, the real number $s$. In our experiments, the EnUKF with different initial ensemble sizes $n=3,4,5,6$ leads to the same value $\text{ceil}(2\bar{l}+1) =13$, so it is not surprising to find in Figs. \ref{fig:enTf_RRmse_vs_delta_meanDoubleL} and \ref{fig:enTf_rmsRatio_vs_delta_meanDoubleL} that the relative rms errors and ratios of the ETKF, which correspond to the EnUKF starting with different initial ensemble sizes, actually coincide. 

From Fig. \ref{fig:enTf_RRmse_vs_delta_meanDoubleL}, one can see that, starting from $\delta=0$, as the covariance inflation factor increases, the relative rmse of the ETKF tends to decrease. However, unlike the situation in the EnUKF, in the test range, as $\delta$ gets larger, say $\delta>7$, the corresponding relative rmse enters a plateau region. The rms ratio indicates a similar behaviour. As $\delta$ increases, the change of the rms ratio becomes smaller, or in other words, the curve appears more and more flat. In order to make the ensemble in the ETKF indistinguishable from the truth, one needs the inflation factor $\delta \approx 1.5$. Like the EnUKF, modest overestimation of the error covariance (i.e., $\delta >1.5$) can also benefit the performance of the ETKF in terms of the relative error.

We use the minimum relative rms errors of the EnUKF and the ETKF to compare their performances. To this end, in Table \ref{tab:min rmse} we list the minimum relative rms errors of the EnUKF with different initial ensemble sizes, and the minimum relative rms errors of the ETKF with the ensemble size about twice the average truncation number plus 1\footnote{Note that different initial ensemble sizes $n=3,4,5,6$ in the EnUKF lead to the same ensemble size in the ETKF. Therefore, in Table \ref{tab:min rmse}, the ETKF has the same minimum relative rmse in different rows.}. Moreover, as a reference, we also show the average noise level (relative rmse) in the observations.

From Table \ref{tab:min rmse}, one can see that the average noise level in the trajectory consisting of the observations is roughly 0.226 (22.6\%), while the minimum relative rmse (i.e. noise level) of the assimilated trajectory through the ETKF scheme is about 0.207 (20.7\%), a reduction of 1.9\% noise level compared to the observations. Similarly, the minimum relative rmse of the assimilated trajectory through the EnUKF scheme is roughly less than 0.175 (17.5\%), a reduction of more than 5\% noise level compared to the observations, and more than 3\% compared to the ETKF scheme. In this sense, both the ETKF and the EnUKF do not diverge as their minimum relative rms errors are less than the average noise level in the observations, yet the EnUKF exhibits a better performance in state estimation than the ETKF.

\subsection{ Effects of parameters on the performance of the EnUKF}

There is a set of adjustable parameters in the EnUKF, e.g.,  threshold $h_1$, lower bound $l_l$ and upper bound $l_u$, $\lambda$ in Eqs. (\ref{enukf: sigma point generation}) and (\ref{enukf:weights}), and $\beta$ in Eqs. (\ref{enukf cov}) and (\ref{enukf background covariances}). In this section we study the effects of these parameters on the performance of the EnUKF. Note that in the previous section, we have already examined the effect of the inflation factor on the performance of the EnUKF. Thus in the subsequent experiments, we will just fix the inflation factor at a particular value (say, zero, but other choice will also do) for the sake of simplicity.

\subsubsection{Effects of the threshold $h_1$ and the bounds $l_l$, $l_u$}

The parameters $h_1$, $l_l$ and $l_u$ determine the truncation numbers $l_k$'s. To examine their effects on the performance of the EnUKF, we fix the covariance inflation factor $\delta=0$, the length scale $l_c=240$, parameters $\lambda=-2$ and $\beta=2$. We specify the upper bound $l_u=6$ in the experiments, but vary the lower bound $l_l$ from $3$ to $6$. This choice is used to represent the typical situation in data assimilation, wherein the ensemble member is often much lower than the dimension of the dynamical system. We also vary the threshold $h_1$, in the logarithmic scale $\log_{10}h_1$, from $2$ to $5.5$, with an even increment of $0.5$ each time. This range represents the moderate values of $h_1$ so as to make the truncation numbers $l_k$ neither too large nor too small. In all the experiments, we initialize the EnUKF with the same initial conditions and background ensemble, with the ensemble size $n=4$ at the first assimilation cycle. 

\begin{figure*}[htb] 
     \centering
     \includegraphics[width=\textwidth]{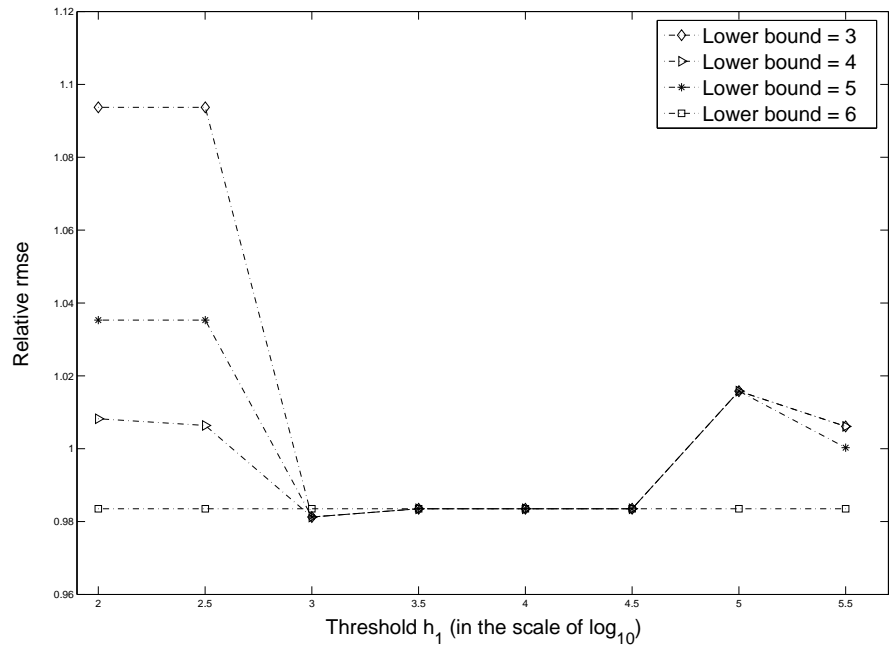}
     \caption{ \label{fig:ukf_RRmse_vs_LOGthreshold} Relative rmse of the EnUKF vs the threshold $h_1$ (in the scale of $\log_{10}$) with different lower bounds $l_l$.}
\end{figure*}

In Fig. \ref{fig:ukf_RRmse_vs_LOGthreshold} we show the simulation results. Clearly, the larger the threshold $h_1$ and the bound $l_l$ are, the larger the truncation number $l_k$ tends to be, which, however, does not mean the better performance in terms of the relative rmse. Indeed, in Fig. \ref{fig:ukf_RRmse_vs_LOGthreshold}, there exists the same optimal threshold 
$\log_{10}h_1=3$ for lower bounds $l_l = 3, 4,5$ \footnote{$l_l=l_u=6$ means $l_k=6$ at every cycle, while the threshold $h_1$ does not affect the choice of $l_k$.}, while the thresholds larger than this value result in larger relative rms errors. For the lower bound $l_l=6$, its relative rms errors are smaller than, or at least equal to those of the bounds $l_l = 3, 4,5$ in most cases. However, at $\log_{10}h_1=3$, the relative rmse given $l_l=6$ is worse than the other cases. To explain these phenomena, we conjecture that, a too small truncation number $l_k$ is not likely to achieve a performance as good as a modest value because it means poor quality of covariance approximation. In contrast, a too large truncation number $l_k$ also does not necessarily achieve a better performance than a modest value because, at some local points of the attractor, a too large truncation number may introduce some spurious structures--from the null space of the SVD-- into the sigma points, which are then treated as equally likely as the other sigma points, and propagated forward to the next cycle. The effect of the spurious structures can be accumulated and eventually deteriorates the overall performance. 

\subsubsection{Effects of the parameters $\lambda$ and $\beta$}\label{sec:lambda-beta}

We proceed to examine the effects of the parameters $\lambda$ and $\beta$. In the experiments, we set the covariance inflation factor $\delta=0$, the length scale $l_c=240$, lower bound $l_l=3$, upper bound $l_u=6$, threshold $h_1=1000$, and initial ensemble size $n=4$. We consider four scenarios with $\beta=0,2,4,6$ respectively. For each value of $\beta$, we compute $20$ values of $\lambda$. To guarantee the positive semi-definiteness of the sample covariances, we start $\lambda$ from $-\beta l_l / (1+\beta)$, with an even increment $\Delta \lambda =1$ each time. In particular, when $\beta=0$ and $\lambda=0$, the effective weight $W_{k,0} + \beta$ of the ensemble mean $\hat{\mathbf{x}}_{k}^a$ equals zero for any $k$. Therefore, in this case, the unscented transform is actually equivalent to the analysis scheme of positive-negative pairs (PNP) in the literature (cf \cite{Wang-which} and the references therein).

\begin{figure*}[htb] 
     \centering
	 \subfigure[Relative rmse vs $\lambda$ with $\beta=0$]{
        \label{fig:ukf_rmse_vs_lambda_beta0}
        \includegraphics[width=0.45\textwidth]{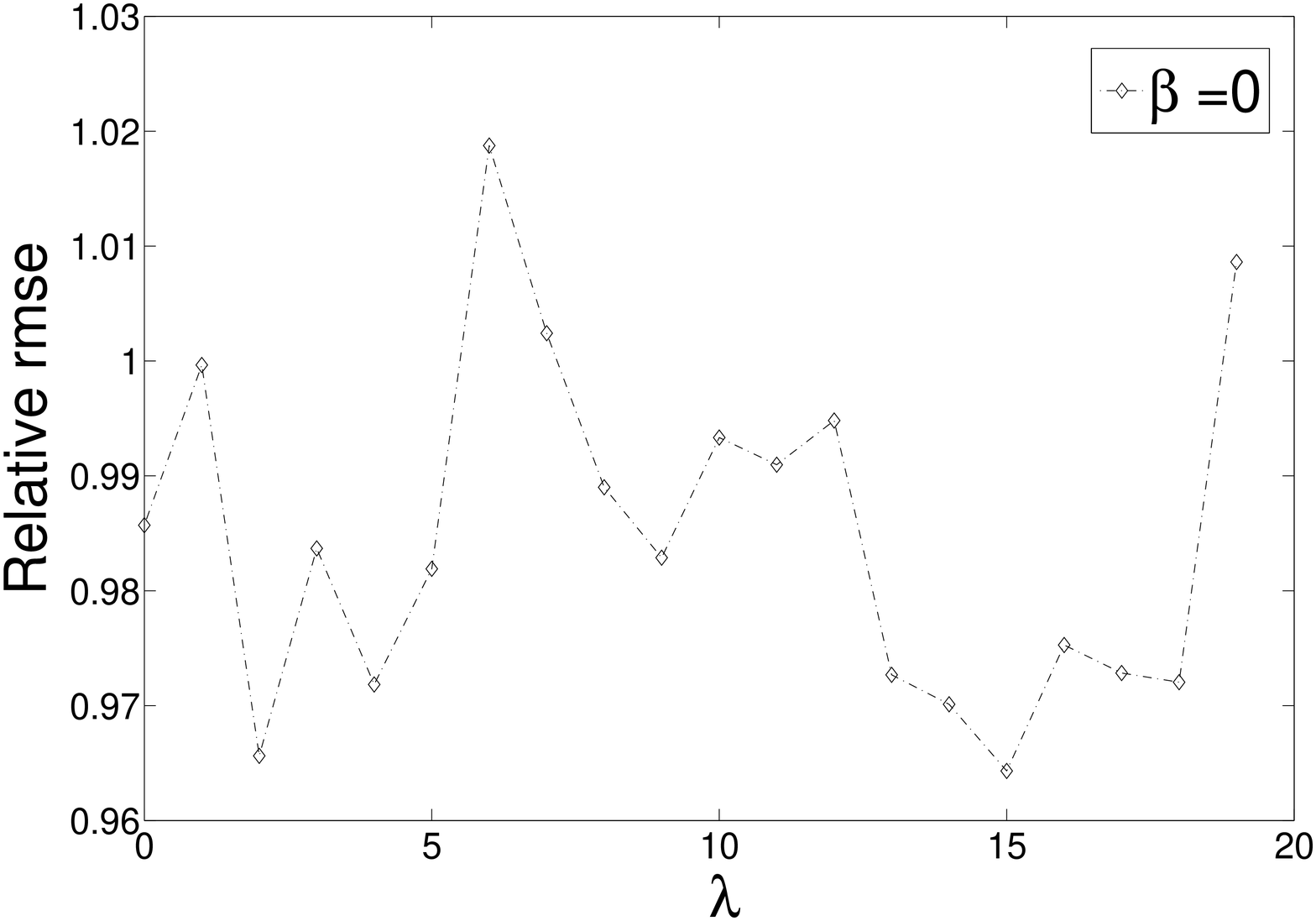}} 
	 \hspace{.3in}
     \subfigure[Relative rmse vs $\lambda$ with $\beta=2$]{
          \label{fig:ukf_rmse_vs_lambda_beta2}
          \includegraphics[width=0.45\textwidth]{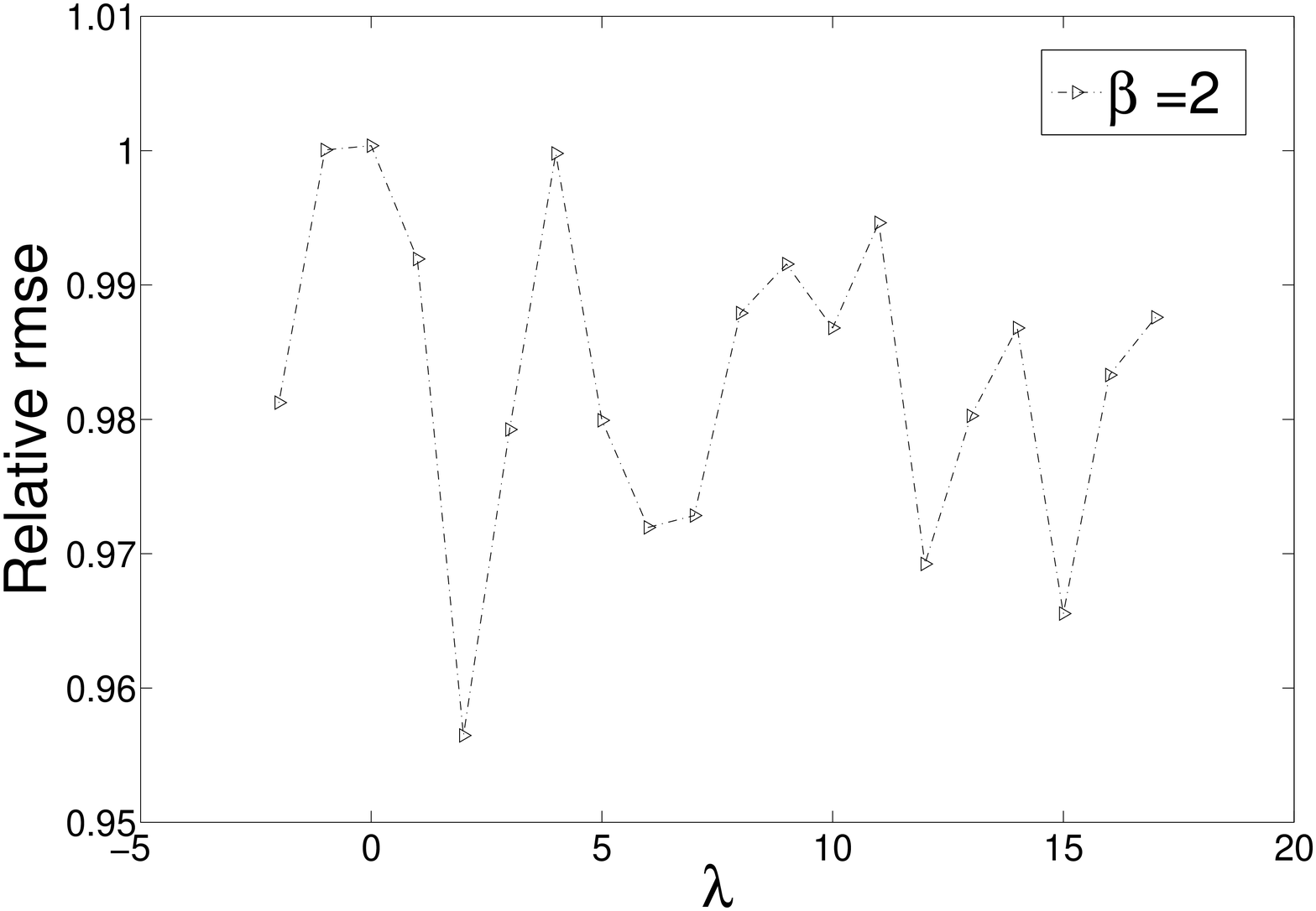}}
    \hspace{.3in}
	\subfigure[Relative rmse vs $\lambda$ with $\beta=4$]{
        \label{fig:ukf_rmse_vs_lambda_beta4}
        \includegraphics[width=0.45\textwidth]{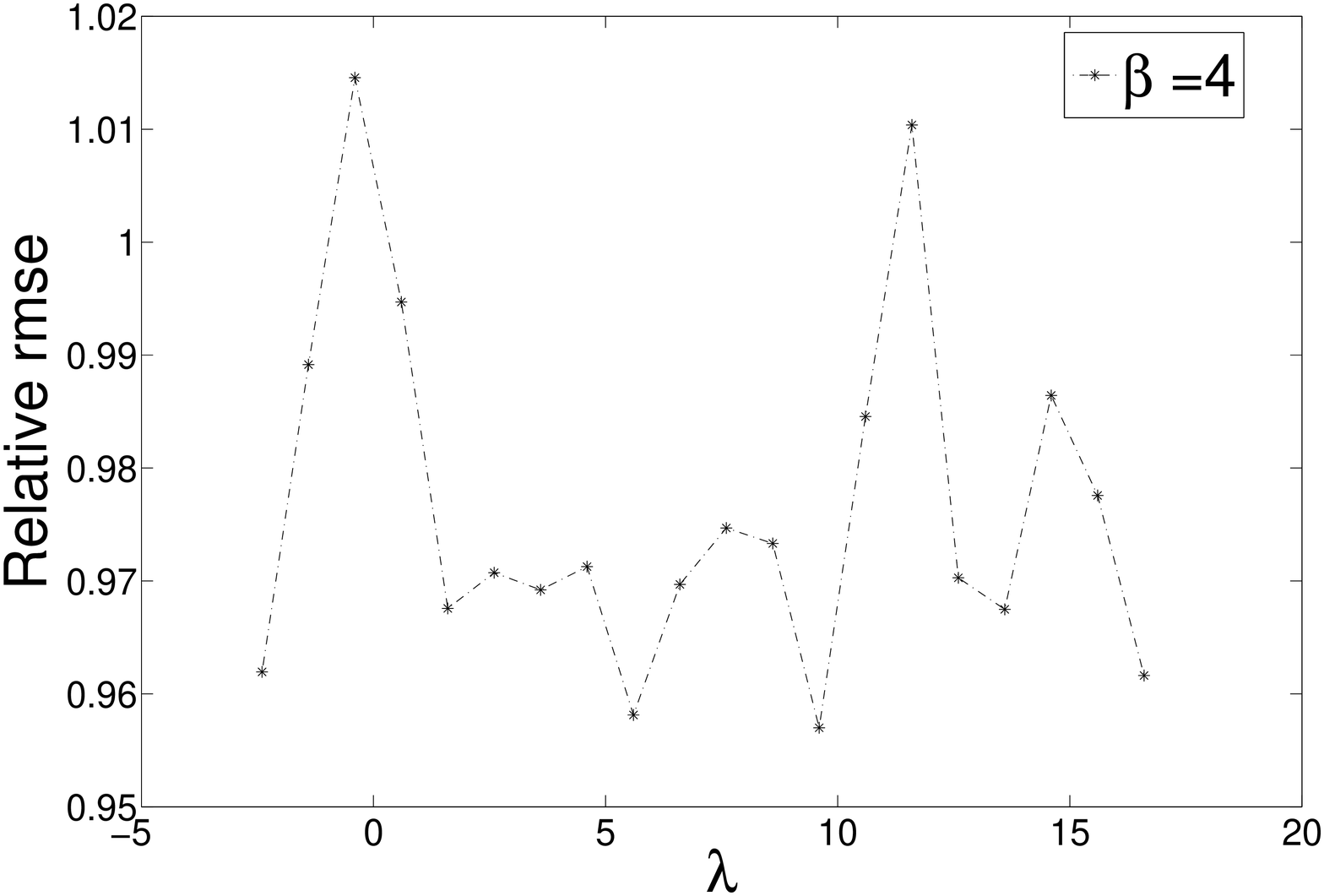}}
	 \hspace{.3in}
     \subfigure[Relative rmse vs $\lambda$ with $\beta=6$]{
          \label{fig:ukf_rmse_vs_lambda_beta6}
          \includegraphics[width=0.45\textwidth]{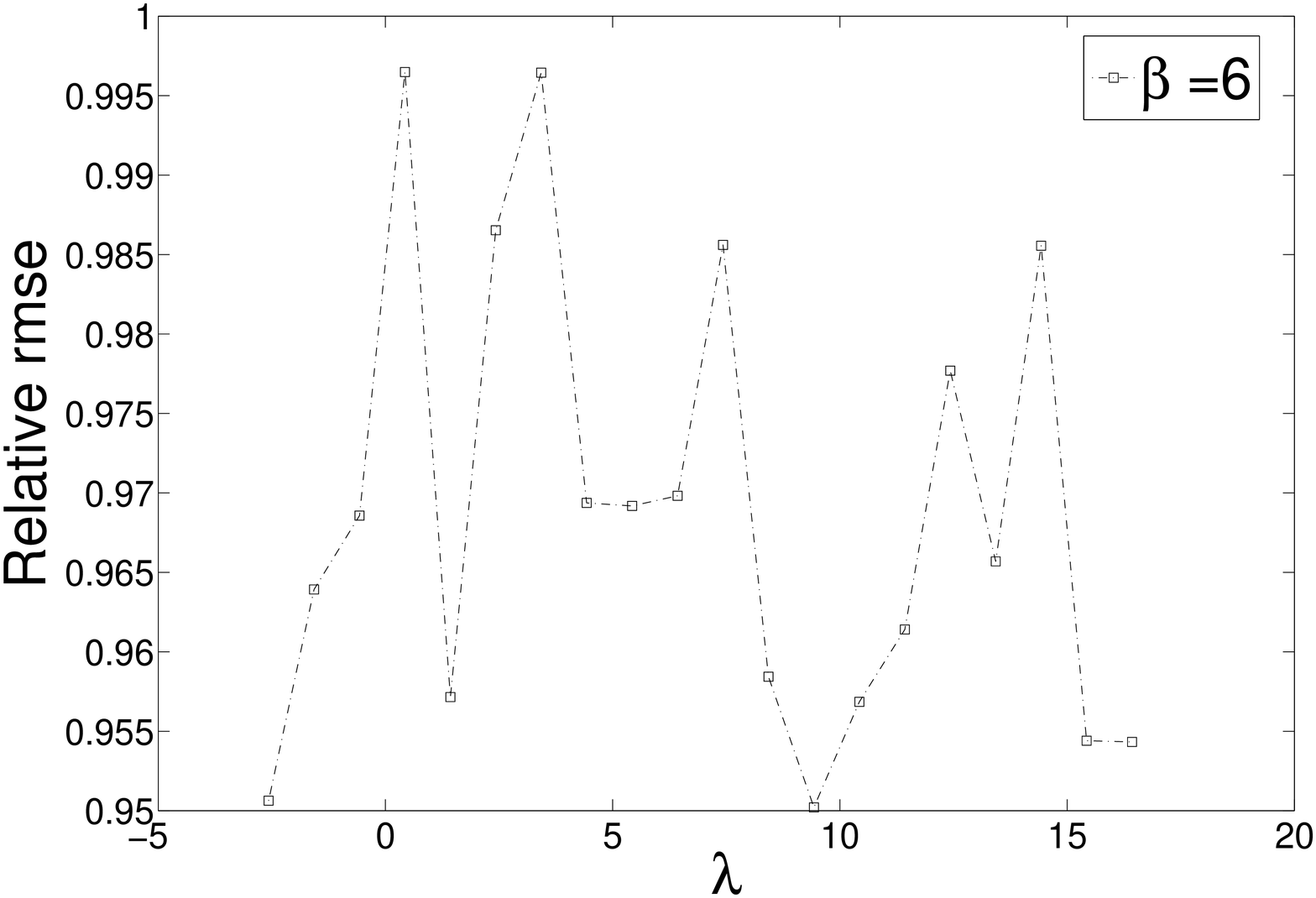}}
     \hspace{.3in}
     \caption{ \label{fig:ukf beta} Effects of the parameters $\beta$ and $\lambda$ on the performance of the EnUKF.}
\end{figure*}

We plot the simulation results in Fig. \ref{fig:ukf beta}. As can be seen, when $\beta$ increases from $0$ to $6$, the minimum relative rmse for a given value of $\beta$ intends to decrease. This may be interpreted as follows: In Eqs. (\ref{enukf cov}) and (\ref{enukf background covariances}), the second terms on the rhs also act like a covariance inflation technique. Therefore, similar to the covariance inflation factor $\delta$, a larger value of $\beta$ tends to result in a smaller minimum relative rmse. 

However, for each fixed $\beta$, there is no clear trend of the optimal value of the parameter $\lambda$. A larger value of $\lambda$ does not imply a smaller relative rmse, or vice verse. Particularly, from Fig. \ref{fig:ukf_rmse_vs_lambda_beta0} it can be seen that, by choosing suitable values for the parameters $\lambda$ and $\beta$, the EnUKF can outperform the EnKF equipped with the analysis scheme of positive-negative pairs. 

As an explanation of the above phenomenon, one may note that, with the other parameters fixed, $\lambda$ is the parameter that determines the relative weights between the sample mean and the other sigma points (cf. Eqs.~(\ref{enukf: sigma point generation}) and (\ref{enukf:weights})). If the underlying system state is linear, then in principle one shall be able to compute the optimal relative weights between the sample mean and the other sigma points under the Gaussianity assumption, and thus determine the optimal value of $\lambda$. Nevertheless, the existence of nonlinearity may make the problem intractable. For nonlinear systems, the optimal relative weights (hence the parameter $\lambda$) may vary from cycle to cycle. However, to search for the optimal parameter $\lambda$ at each assimilation cycle will be computationally expensive, thus in our experiments, we chose to fix the parameter $\lambda$ in the same assimilation window\footnote{Here by ``assimilation window'' we mean the time window from the first assimilation cycle to the maximum.}. This choice cannot reflect the variation of the optimal values of $\lambda$ at different assimilation cycles, therefore it would not be surprising to see that there is a lack of clear trend of the optimal value of $\lambda$ in Fig.~(\ref{fig:ukf_rmse_vs_lambda_beta0}).   

\section{Conclusions}\label{sec:conclusions}

A new ensemble Kalman filter scheme, called the EnUKF, was introduced in this work. The ensemble generation scheme of the EnUKF is similar to the idea of positive-negative pairs (PNP) (\cite{Wang-which} and the references therein), but it differs from the PNP scheme in that, apart from generating symmetric positive-negative pairs, the EnUKF also propagates the ensemble mean forward. Like the EnKF equipped with the PNP scheme, in the EnUKF all symmetric perturbations are associated with an identical weight. Nevertheless, a different weight can be assigned to the ensemble mean. In this sense, the EnUKF can be deemed as a hybrid of central forecast (i.e., the forecast made by propagating the ensemble mean solely) and ensemble forecast (i.e., the forecast made by propagating the ensemble forward), while the PNP scheme is a special case of the unscented transform with the weight of the ensemble mean being zero.

From the analytic results in the Appendix, it can be seen that, in estimation of the sample mean and the covariance of a transformed random variable, the accuracies of the EnUKF will be at least up to second order, no matter whether the original random variable follows a Gaussian distribution or not. If the original random variable follows a symmetric distribution (not necessarily Gaussian), then the accuracies of the estimations increase to third order. Moreover, additional parameters, such as $\lambda$ and $\beta$, are available in the unscented transform to improve the estimation accuracies by choosing proper values. 

By comparing the Taylor series of the transformation function term by term in the Appendix, we also show that the EnUKF has better accuracies than the ordinary EnKF. For numerical verification, using the LE98 model, we compared the performances of the EnUKF and the ETKF in terms of the relative rms errors. Experiment results confirmed that incorporating the unscented transform into an EnKF can benefit its performance. 

\section*{Acknowledgments}
The authors would like to thank Dr Sarah~L. Dance and two anonymous reviewers for their very constructive comments and suggestions.

\clearpage
\section*{ Appendix: Accuracies of the sample mean and covariance of the EnKF and the unscented transform}

In this Appendix we analyze the accuracies of the ordinary EnKF and the EnUKF in estimating the mean and covariance of a random variable transformed by a nonlinear function. In analysis, we assume that the nonlinear function can be expanded in a Taylor series, which converges to the true value of the transformation \cite{Julier2000,Julier2004}.

\subsection*{The actual mean and covariance of the transformed variable in terms of Taylor series}
Given a vector $\bar{\mathbf{z}}$ and a Gaussian perturbation  $\delta \mathbf{z}$ with zero mean and covariance $\mathbf{P}_{zz}$, let us first expand the transform $\mathbf{y} = \mathbf{f} (\bar{\mathbf{z}}+\delta \mathbf{z})$ in a Taylor series around the point $\bar{\mathbf{z}}$, then the mean $\bar{\mathbf{y}}$ is given by
\begin{equation}\label{series expectation}
\begin{split}
\bar{\mathbf{y}} &= \mathbb{E} \left( \mathbf{f} (\bar{\mathbf{z}}+\delta \mathbf{z}) \right) \\
\quad &=\mathbf{f} (\bar{\mathbf{z}}) +  \left( \frac{ \nabla^T \mathbf{P}_{zz}  \nabla }{2!}  \right) \mathbf{f} + \mathbb{E} \left( \frac{\mathbf{D}_{\delta \mathbf{z}}^4 \mathbf{f}}{4!}  + \dotsb \right),
\end{split}
\end{equation}
where the operator
\begin{equation} \label{D def}
\mathbf{D}_{\delta \mathbf{z}} \equiv \delta \mathbf{z}^T \nabla .
\end{equation}

Similarly, the covariance matrix $\mathbf{P}_{yy}$ is given by
\begin{equation} \label{expanded cov of y}
\begin{split}
\mathbf{P}_{yy} =& \mathbb{E} \left( \left( \mathbf{y}- \bar{\mathbf{y}}\right)   \left( \mathbf{y}- \bar{\mathbf{y}}\right)^T \right) \\
= &   \mathbf{J} \mathbf{P}_{zz}  \mathbf{J}^T + \mathbb{E} \left[ \frac{ \mathbf{D}_{\delta \mathbf{z}} \mathbf{f} \left( \mathbf{D}_{\delta \mathbf{z}}^3 \mathbf{f} \right)^T}{3!}   \right. \\
&  +  \frac{ \mathbf{D}_{\delta \mathbf{z}}^2 \mathbf{f} \left( \mathbf{D}_{\delta \mathbf{z}}^2 \mathbf{f} \right)^T}{2! \times 2!} \left.  +  \frac{ \mathbf{D}_{\delta \mathbf{z}}^3 \mathbf{f} \left( \mathbf{D}_{\delta \mathbf{z}} \mathbf{f} \right)^T}{3!} \right]    \\
& - \left[  \left( \frac{ \nabla^T \mathbf{P}_{zz}  \nabla }{2!}
\right) \mathbf{f} \right] \left[  \left( \frac{ \nabla^T
\mathbf{P}_{zz}  \nabla }{2!}  \right) \mathbf{f} \right]^T + \dotsb,
\end{split}
\end{equation}
where $\mathbf{J}=\left( \nabla \mathbf{f} \right)^T \lvert_{\bar{\mathbf{z}}}$ is the Jacobian matrix of $\mathbf{f}$ at $\bar{\mathbf{z}}$.

\subsection*{Accuracies of the EnKF}

In the EnKF, given an ensemble $\{ \mathbf{z}_i\}_{i=1}^n$, the sample mean, in terms of Taylor series, is given by
\begin{equation} \label{enkf series mean}
\begin{split}
\hat{\mathbf{y}} = & \mathbf{f} \left( \bar{\mathbf{z}} \right) +
\frac{\sum_{i=1}^{n}  \mathbf{D}_{\delta \mathbf{z}_i}^2
\mathbf{f}}{n \times 2!} + \frac{\sum_{i=1}^{n}  \mathbf{D}_{\delta
\mathbf{z}_i}^3 \mathbf{f}}{n \times 3!} \\
& + \frac{\sum_{i=1}^{n} \mathbf{D}_{\delta \mathbf{z}_i}^4
\mathbf{f}}{n \times 4!} + \dotsb ,
\end{split}
\end{equation}
where $\delta \mathbf{z}_i = \mathbf{z}_i -  \hat{\mathbf{z}}$ with $  \hat{\mathbf{z}} = \sum_{i=1}^{n} \mathbf{z}_i /n$ being the sample mean of $\mathbf{z}$.

In the rhs of Eq. (\ref{enkf series mean}), 
\[
\frac{\sum_{i=1}^{n}  \mathbf{D}_{\delta \mathbf{z}_i}^2 \mathbf{f}}{n \times 2!} = \frac{1}{2!}  \nabla^T \left( \frac{1}{n} \sum\limits_{i=1}^{n}  \delta \mathbf{z}_i  \delta \mathbf{z}_i^T \right) \nabla  \mathbf{f},
\]
which is a biased estimation of the second term on the rhs of Eq. (\ref{series expectation}). Moreover, spurious modes may rise in higher order terms of Eq. (\ref{enkf series mean}), for example, the third order term
\begin{equation} \label{sample D3}
\frac{1}{n} \sum_{i=1}^{n}  \mathbf{D}_{\delta \mathbf{z}_i}^3 \mathbf{f}= \nabla^T \left( \frac{1}{n} \sum\limits_{i=1}^{n} \delta \mathbf{z}_i  \delta \mathbf{z}_i^T \nabla \delta \mathbf{z}_i^T \right) \nabla \mathbf{f}
\end{equation}
in general will not vanish.

Similarly, we have the sample covariance
\begin{equation} \label{series enkf cov}
\begin{split}
\hat{\mathbf{P}}_{yy} = &  \mathbf{J} \mathbf{P}_{zz}  \mathbf{J}^T + \frac{1}{(n-1) \times 2! } \sum\limits_{i=1}^{n} \left[ \mathbf{D}_{\delta \mathbf{z}_i} \mathbf{f}  \left( \mathbf{D}_{\delta \mathbf{z}_i}^2 \mathbf{f}\right) ^T \right]\\
&+ \frac{1}{n-1} \left( \frac{ \sum_{i=1}^{n}  \mathbf{D}_{\delta \mathbf{z}_i} \mathbf{f}  \left( \mathbf{D}_{\delta \mathbf{z}_i}^3 \mathbf{f}\right) ^T }{3!} + \right. \\
& \left. \frac{ \sum_{i=1}^{n}  \mathbf{D}_{\delta \mathbf{z}_i} \mathbf{f}^2  \left( \mathbf{D}_{\delta \mathbf{z}_i}^2 \mathbf{f}\right) ^T }{ 2! \times 2!}   + \frac{ \sum_{i=1}^{n}  \mathbf{D}_{\delta \mathbf{z}_i}^3 \mathbf{f}  \left( \mathbf{D}_{\delta \mathbf{z}_i} \mathbf{f}\right) ^T }{3!} \right) \\
& - \frac{n-1}{n}  \left[  \left( \frac{ \nabla^T \mathbf{P}_{zz}  \nabla }{2!}  \right) \mathbf{f} \right] \left[  \left( \frac{ \nabla^T \mathbf{P}_{zz}  \nabla }{2!}  \right) \mathbf{f} \right]^T  + \dotsb \, .
\end{split}
\end{equation}
Note that here 
\[
\mathbf{P}_{zz} = \frac{1}{n-1} \sum\limits_{i=1}^{n} \delta \mathbf{z}_i \delta \mathbf{z}_i^T . 
\]

Comparing Eq. (\ref{series enkf cov}) and Eq. (\ref{expanded cov of y}), we note that
\begin{itemize}
\item Compared to Eq. (\ref{expanded cov of y}), there are also spurious modes arising in higher order terms in Eq. (\ref{series enkf cov});
\item There is a bias in the term $\left[  \left( \nabla^T \mathbf{P}_{zz}  \nabla   \right) \mathbf{f} \right] \left[  \left( \nabla^T \mathbf{P}_{zz}  \nabla   \right) \mathbf{f} \right]^T$.
\end{itemize}

\subsection*{Accuracies of the unscented transform}
For the unscented transform, given a set of sigma points $\left \{ \mathcal{Z}_i\right \}_{i=0}^{2L}$ with mean $\bar{\mathbf{z}}$ and covariance $\mathbf{P}_{zz}$, the sample mean, in terms of Taylor series, is given by
\begin{equation} \label{series ut mean}
\begin{split}
\hat{\mathbf{y}} &= \sum\limits_{i=0}^{2L} W_i \mathbf{f} \left( \mathcal{Z}_i \right) \\
& = \mathbf{f} \left(  \bar{\mathbf{z}} \right) + \left( \frac{ \nabla^T \mathbf{P}_{zz}  \nabla }{2!}  \right) \mathbf{f} + \frac{1}{2(L+\lambda)} \sum\limits_{i=1}^{2L} \left( \mathbf{D}_{\delta \mathbf{z}_i}^4 \mathbf{f} + \dotsb \right) .
\end{split}
\end{equation}
Note that in Eq. (\ref{series ut mean}), the third order terms vanish because of the symmetry in the sigma points.

Comparing Eq. (\ref{series ut mean}) with Eq. (\ref{series expectation}), we note that, second and third order terms in Eq. (\ref{series ut mean}) match those in Eq. (\ref{series expectation}) exactly, while the difference starts from fourth order terms.

Similarly, the sample covariance is given by
\begin{equation} \label{series ut cov}
\begin{split}
\hat{\mathbf{P}}_{yy} =& \sum\limits_{i=0}^{2L} W_i \left(  \mathbf{f} \left( \mathcal{Z}_i \right) - \hat{\mathbf{y}} \right) \left(  \mathbf{f} \left( \mathcal{Z}_i \right) - \hat{\mathbf{y}} \right) ^T \\
= &  \mathbf{J} \mathbf{P}_{zz}  \mathbf{J}^T + \frac{1}{2(L+\lambda)} \left( \frac{ \sum_{i=1}^{2L}  \mathbf{D}_{\delta \mathbf{z}_i} \mathbf{f}  \left( \mathbf{D}_{\delta \mathbf{z}_i}^3 \mathbf{f}\right) ^T }{3!} + \right. \\
& \left. \frac{ \sum_{i=1}^{2L}  \mathbf{D}_{\delta \mathbf{z}_i} \mathbf{f}^2  \left( \mathbf{D}_{\delta \mathbf{z}_i}^2 \mathbf{f}\right) ^T }{ 2! \times 2!}   + \frac{ \sum_{i=1}^{2L}  \mathbf{D}_{\delta \mathbf{z}_i}^3 \mathbf{f}  \left( \mathbf{D}_{\delta \mathbf{z}_i} \mathbf{f}\right) ^T }{3!} \right) \\
& - \left[  \left( \frac{ \nabla^T \mathbf{P}_{zz}  \nabla }{2!}  \right) \mathbf{f} \right] \left[  \left( \frac{ \nabla^T \mathbf{P}_{zz}  \nabla }{2!}  \right) \mathbf{f} \right]^T  + \dotsb \, .
\end{split}
\end{equation}

Clearly, unlike Eq. (\ref{series enkf cov}), there is no spurious mode rising in third order terms in the unscented transform. Moreover, there is also no bias in the term 
\[
\left[  \left( \nabla^T \mathbf{P}_{zz}  \nabla   \right) \mathbf{f} \right] \left[  \left( \nabla^T \mathbf{P}_{zz}  \nabla   \right) \mathbf{f} \right]^T .
\]

To further reduce the approximation errors in fourth order terms, one may introduce an additional term, $\beta \left(  \mathbf{f} \left( \mathcal{Z}_0 \right) - \hat{\mathbf{y}} \right) \left(  \mathbf{f} \left( \mathcal{Z}_0 \right) - \hat{\mathbf{y}} \right) ^T$, to the sample covariance such that it can be re-written as
\begin{equation} \tag{$\ref{ut cov}$}
\begin{split}
 &\hat{\mathbf{P}}_{yy} =  \sum\limits_{i=0}^{2L} W_i \, \left( \mathbf{f} \left( \mathcal{Z}_{i} \right) - \hat{\mathbf{y}} \right) \left(  \mathbf{f} \left( \mathcal{Z}_{i} \right) - \hat{\mathbf{y}} \right ) ^T + \beta \left( \mathbf{f} \left( \mathcal{Z}_{0} \right) - \hat{\mathbf{y}} \right) \left(  \mathbf{f} \left( \mathcal{Z}_{0} \right) - \hat{\mathbf{y}} \right ) ^T \, .
\end{split}
\end{equation}
It can be shown that (cf. Eq. (21) of \cite{Julier2004}), in the case that $\mathbf{z}$ follows a Gaussian distribution, choosing $\beta=2$ will minimize the approximation errors in fourth order terms.

\bibliographystyle{elsart-num-sort}
\bibliography{./references}

\begin{thebibliography}{10}
\expandafter\ifx\csname url\endcsname\relax
  \def\url#1{\texttt{#1}}\fi
\expandafter\ifx\csname urlprefix\endcsname\relax\def\urlprefix{URL }\fi

\bibitem{Anderson-ensemble}
J.~L. Anderson, An ensemble adjustment kalman filter for data assimilation,
  Mon. Wea. Rev. 129 (2001) 2884--2903.

\bibitem{Anderson-Monte}
J.~L. Anderson, S.~L. Anderson, A monte carlo implementation of the nonlinear
  filtering problem to produce ensemble assimilations and forecasts, Mon. Wea.
  Rev. 127 (1999) 2741--2758.

\bibitem{Beezley-morphing}
J.~D. Beezley, J.~Mandel, Morphing ensemble kalman filters, Tellus 60A (2007)
  131 -- 140.

\bibitem{Bishop-adaptive}
C.~H. Bishop, B.~J. Etherton, S.~J. Majumdar, Adaptive sampling with ensemble
  transform kalman filter. part {I}: theoretical aspects, Mon. Wea. Rev. 129
  (2001) 420--436.

\bibitem{Burgers-analysis}
G.~Burgers, P.~J. van Leeuwen, G.~Evensen, On the analysis scheme in the
  ensemble kalman filter, Mon. Wea. Rev. 126 (1998) 1719--1724.

\bibitem{Ehrendorfer1997}
M.~Ehrendorfer, J.~J. Tribbia, Optimal prediction of forecast error covariances
  through singular vectors, J. Atmos. Sci. 54 (1997) 286--312.

\bibitem{Evensen-using}
G.~Evensen, Using the extended kalman filter with a multilayer
  quasi-geostrophic ocean model, J. Geophys. Res. 97 (1992) 17,905--17,924.

\bibitem{Evensen-sequential}
G.~Evensen, Sequential data assimilation with a nonlinear quasi-geostrophic
  model using monte carlo methods to forecast error statistics, J. Geophys.
  Res. 99(C5) (1994) 10,143--10,162.

\bibitem{Evensen-assimilation}
G.~Evensen, P.~J. van Leewen, Assimilation of geosat altimeter data for the
  aghulas current using the ensemble kalman filter with a quasi-geostrophic
  model, Mon. Wea. Rev. 124 (1996) 85--96.

\bibitem{Golub-matrix}
G.~H. Golub, C.~F. Van~Loan, Matrix Computations, JHU Press, 1996.

\bibitem{Hamill-distance}
T.~M. Hamill, J.~S. Whitaker, C.~Snyder, Distance-dependent filtering of
  background error covariance estimates in an ensemble kalman filter, Mon. Wea.
  Rev. 129 (2001) 2776--2790.

\bibitem{Han-evaluation}
X.~Han, X.~Li, An evaluation of the nonlinear/non-{G}aussian filters for the
  sequential data assimilation, Remote Sensing of Environment 112 (2008) 1434
  -- 1449.

\bibitem{Hansen-truncated}
P.~C. Hansen, The truncated svd as a method for regularization, BIT 27 (1987)
  534 -- 553.

\bibitem{Houtekamer1998}
P.~L. Houtekamer, H.~L. Mitchell, Data assimilation using an ensemble kalman
  filter technique, Mon. Wea. Rev. 126 (1998) 796--811.

\bibitem{Houtekamer-sequential}
P.~L. Houtekamer, H.~L. Mitchell, A sequential ensemble kalman filter for
  atmospheric data assimilation, Mon. Wea. Rev. 129 (2001) 123--137.

\bibitem{Julier2000}
S.~Julier, J.~Uhlmann, H.~Durrant-Whyte, A new method for the nonlinear
  transformation of means and covariances in filters and estimators, IEEE
  Transactions on Automatic Control 45 (2000) 477--482.

\bibitem{Julier2004}
S.~J. Julier, J.~K. Uhlmann, Unscented filtering and nonlinear estimation,
  Proc. IEEE 92 (2004) 401--422.

\bibitem{Kalnay-atmospheric}
E.~Kalnay, Atmospheric Modeling, Data Assimilation and Predictability,
  Cambridge University Press, 2002.

\bibitem{Kalnay-4dvar}
E.~Kalnay, H.~Li, T.~Miyoshi, S.-C. Yang, J.~Ballabrera-Poy, 4-{D}-var or
  ensemble kalman filter, Tellus 59A (2007) 758--773.

\bibitem{Livings-unbiased}
D.~M. Livings, S.~L. Dance, N.~K. Nichols, Unbiased ensemble square root
  filters, Physica D 237 (2008) 1021 -- 1028.

\bibitem{Lorenz-predictability}
E.~N. Lorenz, Predictability-a problem solved, in: T.~Palmer (ed.),
  Predictability, ECMWF, Reading, UK, 1996.

\bibitem{Lorenz-optimal}
E.~N. Lorenz, K.~A. Emanuel, Optimal sites for supplementary weather
  observations: Simulation with a small model, J. Atmos. Sci. 55 (1998)
  399--414.

\bibitem{Maybeck-stochastic}
P.~Maybeck, Stochastic Models, Estimation, and Control, Academic Press, 1979.

\bibitem{Murphy-impact}
J.~M. Murphy, The impact of ensemble forecasts on predictability, Quart. J.
  Roy. Meteor. Soc. 114 (1988) 463 -- 493.

\bibitem{Ott-local}
E.~Ott, B.~R. Hunt, I.~Szunyogh, A.~V. Zimin, E.~J. Kostelich, E.~J. Corazza,
  E.~Kalnay, D.~J. Patil, J.~A. Yorke, A local ensemble kalman filter for
  atmospheric data assimilation, Tellus 56A (2004) 415--428.

\bibitem{Sakov2007}
P.~Sakov, P.~R. Oke, A deterministic formulation of the ensemble kalman filter:
  an alternative to ensemble square root filters, Tellus 60A (2008) 361--171.

\bibitem{Smith-cluster}
K.~W. Smith, Cluster ensemble {K}alman filter, Tellus 59A (2007) 749--757.

\bibitem{Tippett-ensemble}
M.~K. Tippett, J.~L. Anderson, C.~H. Bishop, T.~M. Hamill, J.~S. Whitaker,
  Ensemble square root filters, Mon. Wea. Rev. 131 (2003) 1485--1490.

\bibitem{Treebushny-construction}
D.~Treebushny, H.~Madsen, On the construction of a reduced rank square-root
  kalman filter for efficient uncertainty propagation, Future Generation
  Computer Systems 21 (2005) 1047 -- 1055.

\bibitem{Turner-ensemble}
M.~R.~J. Turner, J.~P. Walker, P.~R. Oke, Ensemble member generation for
  sequential data assimilation, Remote Sensing of Environment 112 (2008) 1421
  -- 1433.

\bibitem{Uzunoglu-adaptive}
B.~Uzunoglu, S.~J. Fletcher, Z.~M., I.~M. Navon, Adaptive ensemble reduction
  and inflation, Quart. J. Roy. Meteor. Soc. 133 (2005) 1281 -- 1294.

\bibitem{vanLeeuwen-variance}
P.~J. van Leeuwen, A variance minimizing filter for large-scale applications,
  Mon. Wea. Rev. 131 (2003) 2071--2084.

\bibitem{Wang-which}
X.~Wang, C.~H. Bishop, S.~J. Julier, Which is better, an ensemble of
  positive-negative pairs or a centered simplex ensemble, Mon. Wea. Rev. 132
  (2004) 1590--1605.

\bibitem{Whitaker-ensemble}
J.~S. Whitaker, T.~M. Hamill, Ensemble data assimilation without perturbed
  observations, Mon. Wea. Rev. 130 (2002) 1913--1924.

\bibitem{Zupanski-maximum}
M.~Zupanski, Maximum likelihood ensemble filter: theoretical aspects, Mon. Wea.
  Rev. 133 (2005) 1710--1726.

\end{thebibliography}

\clearpage

\begin{table*}[htb] 
\caption{\label{tab:min rmse} Minima of the relative rms errors in Figs. \ref{fig:ukf_RRmse_VS_InflationFactor_lownEn} and \ref{fig:enTf_RRmse_vs_delta_meanDoubleL} and the average relative rmse (noise level) $e^{obv}_r$ of the observations.}
\begin{tabular}{c c c c}
\\
\hline \hline
Initial ensemble size & EnUKF & ETKF  & Observations \\ 
\hline
$n=3$  &0.1719 & 0.2074 & \multirow{4}{*}{0.2256}\\
$n=4$ &0.1722 & 0.2074 \\
$n=5$ & 0.1730 & 0.2074 \\
$n=6$ &  0.1753 & 0.2074 \\
\hline \hline
\end{tabular}
\end{table*}
\clearpage

\clearpage

\clearpage

\clearpage

\clearpage

\end{document}